\documentclass[11pt,letterpaper]{article}
\usepackage[in]{fullpage}
\usepackage{amsmath,amsthm,amsfonts,amssymb}
\usepackage{liyang}

\usepackage{framed}
\usepackage{verbatim}
\usepackage{enumitem}
\usepackage{array}
\usepackage{multirow}
\usepackage{afterpage}

\usepackage{caption}

\usepackage[usenames,dvipsnames]{xcolor}

\makeatletter
\newtheorem*{rep@theorem}{\rep@title}
\newcommand{\newreptheorem}[2]{
\newenvironment{rep#1}[1]{
 \def\rep@title{#2 \ref{##1}}
 \begin{rep@theorem}\itshape}
 {\end{rep@theorem}}}
\makeatother
\theoremstyle{plain}

\def\colorful{1}

\ifnum\colorful=1

\fi
\ifnum\colorful=0

\fi

\usepackage{boxedminipage}

\newcommand{\ignore}[1]{}

\newreptheorem{theorem}{Theorem}
\newtheorem*{theorem*}{Theorem}
\newreptheorem{lemma}{Lemma}
\newreptheorem{proposition}{Proposition}
\newtheorem*{noclaim*}{Claim}

\title{Boolean Unateness Testing with $\widetilde{O}(n^{3/4})$ Adaptive Queries\vspace{0.6cm}}

\author{
Xi Chen\thanks{Columbia University, email: \texttt{xichen@cs.columbia.edu}.}
\and
Erik Waingarten\thanks{Columbia University, email: \texttt{eaw@cs.columbia.edu}.}
\and
Jinyu Xie\thanks{Columbia University, email: \texttt{jinyu@cs.columbia.edu}}}
\begin{document}
\maketitle
\thispagestyle{empty}

\begin{abstract}
We give an adaptive algorithm which tests whether an unknown Boolean \mbox{function} $f\colon \{0, 1\}^n \to \{0, 1\}$ is unate, i.e. every variable of $f$ is either non-decreasing or non-increasing,~or $\eps$-far from unate with one-sided error using $\smash{\widetilde{O}(n^{3/4}/\eps^2)}$ queries. This improves on the best adaptive $O(n/\eps)$-query~algorithm from Baleshzar, Chakrabarty, Pallavoor, Raskhodnikova and Seshadhri \cite{BCPRS17} when $1/\eps\ll n^{1/4}$. Combined with the 
$\smash{\widetilde{\Omega}(n)}$-query lower~bound~for  non-adaptive algorithms with one-sided error of \cite{CWX17, BCPRS17b}, we conclude that adaptivity helps for the testing of unateness with one-sided error. A crucial component of our algorithm 
  is a new subroutine for finding bi-chromatic edges in the Boolean hypercube called \emph{adaptive edge search}.

\end{abstract}\newpage

\hypersetup{linkcolor=magenta}
\hypersetup{linktocpage}
\setcounter{tocdepth}{2}
\setcounter{totalnumber}{1}



\setcounter{page}{1}

\def\bi{\boldsymbol{i}}
\def\score{\textsc{Score}}
\def\wto{\widetilde{O}} \def\wtomega{\widetilde{\Omega}}
\def\rsb{\textsc{Rand-Search}} \def\good{\textsc{Good-Set}}
\def\goodf{\textsc{Good-Frac}}
\def\ESearch{\textsc{AE-Search}}


\section{Introduction}

A Boolean function $f \colon \{0, 1\}^n \to \{0, 1\}$ is \emph{monotone}
  if every variable of $f$ is non-decreasing, and is
  \emph{unate} if every variable of $f$ is either non-decreasing or non-increasing
  (or equivalently, there exists an $r\in \{0,1\}^n$ such that $g(x)=f(x\oplus r)$ is monotone,
  where $\oplus$ is the bit-wise XOR).~Both~problems of testing \emph{monotonicity}
  and \emph{unateness}
  were first introduced in \cite{GGLRS00}.
The goal is to design~an algorithm that decides whether
  an unknown $f:\{0,1\}^n\rightarrow \{0,1\}$ has the property being tested~or is far from having the property (see Section \ref{sec:preliminaries} for the formal definition) with as few queries as possible.
After a sequence~of developments from the past few years
  \cite{CS14, CST14, KMS15, CDST15, BB16, CS16, CWX17}, the query complexity of non-adaptive
  algorithms for monotonicity has been pinned down at $\smash{\widetilde{\Theta}(n^{1/2})}$;
  for adaptive monotonicity testing algorithms
  there remains a gap between $\widetilde{O}(n^{1/2})$
  and $\widetilde{\Omega}(n^{1/3})$. 
The query complexity of testing unateness, however, is less well-understood.

The seminal work of \cite{GGLRS00} presented an $O(n^{3/2}/\eps)$-query algorithm
  for testing unateness of Boolean functions.
It proceeds by sampling $O(n^{3/2}/\eps)$ \emph{edges}\footnote{A pair of points $(x,y)$ in
$\{0,1\}^n$ is
  an edge in the Boolean hypercube if $x_i\ne y_i$ at exactly one coordinate $i\in [n]$.
We will refer to $i$ as the direction~of $(x,y)$.
An edge $(x,y)$
  along direction $i$ is \emph{bi-chromatic}
  (in $f$) if $f(x)\ne f(y)$; it~is \emph{monotone} if it is bi-chromatic and has
  $f(x)=x_i$;
  it is \emph{anti-monotone} if it is bi-chromatic but not~monotone.}
  of $\{0,1\}^n$ uniformly at random, 
  and rejects only when it finds a so-called \emph{edge violation} ---
  two edges $(x, y)$ and $(x', y')$ in the same direction $i$ for some $i\in [n]$ such that
  one is monotone and the other is anti-monotone.~By definition 
the existence of an edge violation ensures that the function is not unate~and thus,
  the algorithm has one-sided error (i.e., it always accepts a unate function); we refer to algorithms which always accept unate functions as \emph{one-sided}.
This algorithm is non-adaptive as well (i.e.,~all queries can be made at once).
For the correctness, \cite{GGLRS00} showed that after sampling $O(n^{3/2}/\eps)$ random edges an edge violation is found with high probability when $f$ is $\eps$-far from unate.


Recently, \cite{KS16} obtained the first improvement to the upper bound of \cite{GGLRS00} by presenting an
  ${O}( n\log n/\eps)$-query  adaptive, one-sided algorithm.
Later, \cite{BMPR16} generalized the algorithm~to work for real-valued functions over the $n$-dimensional hypergrid, $f \colon [m]^n \to \R$.
The current best upper~bounds for testing unateness of Boolean functions are $O((n/\eps)\log(n/\eps))$ for non-adaptive algorithms
  \cite{CS16b,BCPRS17}, and $O(n/\eps)$ for adaptive algorithms \cite{BCPRS17} (with a logarithmic
  advantage). Both algorithms work for
  real-valued functions 
  and are shown to be optimal for real-valued functions  in \cite{BCPRS17}.

On the lower bound side, \cite{BMPR16} was the first to give a lower bound on testing unateness by showing that any non-adaptive algorithm with one-sided error must make $\Omega(\sqrt{n}/\eps)$ many queries. Then, \cite{CWX17} showed that unateness testing of Boolean functions
  requires $\widetilde{\Omega}(n^{2/3})$ queries for adaptive algorithms with two-sided error, showing that a polynomial gap between testing monotonicity and unateness for Boolean functions.\footnote{The conference version of the paper included a weaker lower bound of $\widetilde{\Omega}(\sqrt{n})$ for testing unateness. Since then, the authors have improved the lower bound to $\widetilde{\Omega}(n^{2/3})$ and have updated the full-version of the paper, available as \texttt{arXiv:1702.06997}.} For non-adaptive algorithms with one-sided error, \cite{CWX17, BCPRS17b} show $\widetilde{\Omega}(n)$ queries are necessary (for some constant $\eps>0$), which shows the~\mbox{algorithm} of \cite{CS16b,BCPRS17} is optimal among non-adaptive algorithms with one-sided error for Boolean functions. \vspace{-0.28cm}

\paragraph{Our Contribution.}  Generally, the power of adaptivity in property testing of Boolean functions
  is not yet well understood. 
Taking the examples of monotonicity and unateness,
  the current best algorithms are both non-adaptive\hspace{0.03cm}\footnote{For real-valued functions,
  \cite{BCPRS17} showed that adaptivity helps by a logarithmic factor.} (ignoring polylogarithmic factors),
  and polynomial gaps remain between the best upper and lower bounds for the query complexity of adaptive algorithms.



The main result of this work is an $\smash{\widetilde{O}(n^{3/4}/\eps^2)}$-query adaptive, one-sided algorithm for unateness
  testing of Boolean functions.

\begin{theorem}[Main]
\label{thm:main}
There is an $\widetilde{O}(n^{3/4}/\eps^2)$-query\hspace{0.01cm},\footnote{See (\ref{hehewiwi}) for the hidden
  polylogarithmic factor; we have made no effort to optimize the polynomial dependence on $\log n$ and $\log(1/\eps)$ in the algorithm.} adaptive algorithm
  with the following property: Given an 
$\eps>0$ and query access
to an unknown $f\colon \{0,1\}^n \to \{0,1\}$,
  it always returns ``unate''~if~$f$ is unate and 
  returns ``non-unate'' with probability at least $2/3$ if $f$ is $\eps$-far from unate.
\end{theorem}

Compared to the $\widetilde{\Omega}(n)$ lower bound for non-adaptive, one-sided algorithms~\cite{CWX17}, Theorem~\ref{thm:main} implies that adaptivity helps by a polynomial factor for one-sided algorithms.
Additionally, given the lower bound of $\Omega(n/\eps)$ for unateness testing of real-valued functions
  over $\{0,1\}^n$ \cite{BCPRS17},~our result shows that Boolean functions are polynomially easier to test than real-valued functions. The current known upper and lower bounds for testing unateness for $\eps = \Theta(1)$ are summarized in the Table~1. 
  
\begin{table}\label{tab:current}
\centering
\begin{tabular}{|c||c|c|}
\hline
 & Adaptive & Non-adaptive \\
 \hline \hline
 Upper bounds & $\widetilde{O}(n^{3/4})$ (this work) & $O(n)$ \cite{BCPRS17}\\
 \hline
 Lower bounds & $\widetilde{\Omega}(n^{2/3})$ \cite{CWX17} & $\widetilde{\Omega}(n)$ (one-sided) \cite{CWX17, BCPRS17b} \\
 \hline
\end{tabular}
\caption{Current knowledge on upper and lower bounds for testing unateness. We consider the regime where $\eps = \Theta(1)$.}
\end{table}

Our algorithm is heavily inspired by the work of \cite{KMS15}, where they
  prove a  directed analogue of an isoperimetric inequality of Talagrand \cite{T93}
  and used it to reveal strong connections between the structure of
  anti-monotone edges of a Boolean function and its distance to monotonicity.~In~particular, their inequality implies that when~$f$~is~far from monotone, there must exist
  a  highly regular bipartite graph of certain size that consists of anti-monotone edges of $f$ only (see Theorem \ref{thm:kms}).
The analysis of~our~algorithm relies on this implication. (See more discussion later in
  Section \ref{sketchsec}.)

A recent work of \cite{CG17} introduced the notion of
  ``\emph{rounds}'' of adaptivity to quantify the degree of adaptivity used by a property testing algorithm.
We notice that our algorithm can be implemented using only two rounds of adaptivity.\vspace{-0.1cm}

\subsection{Binary search versus adaptive edge search\vspace{-0.05cm}}\label{sketchsec}

We give some high-level ideas behind our main algorithm.
First, it outputs ``non-unate'' only when an edge violation is found and thus,
  it is one-sided. Our analysis focuses on showing that, given
  a function that is $\eps$-far from unate, the algorithm finds an edge violation with high probability.

\begin{figure}
\centering
\begin{picture}(230,230)
    \put(0,0){\includegraphics[width=0.5\linewidth]{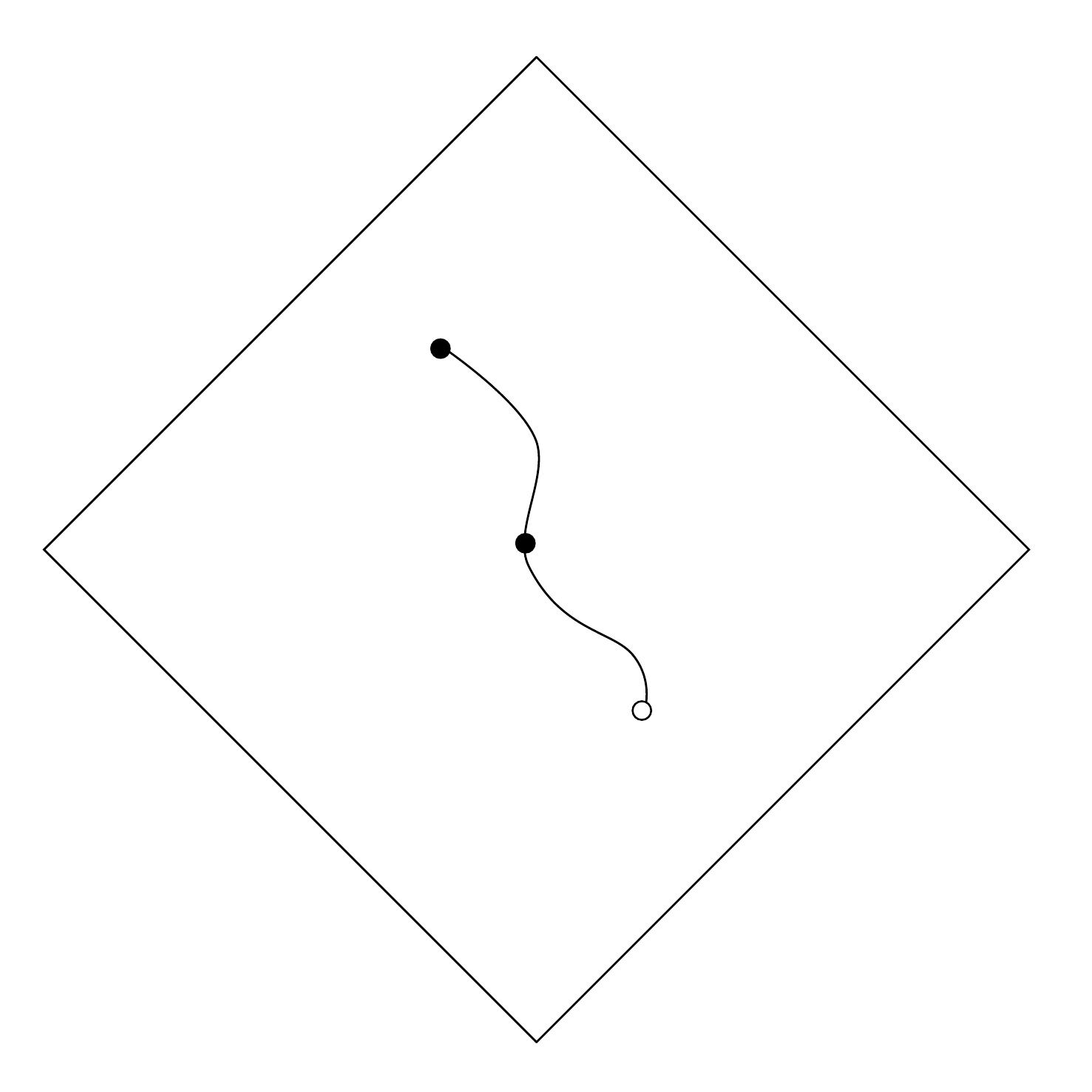}}
    \put(103, 115){$z$}
    \put(85, 150){$y$}
    \put(130, 74){$x$}
  \end{picture}
\caption{Pictorial representation of one step of the binary search strategy for finding bi-chromatic edges. The hypercube $\{0, 1\}^n$ is represented as the diamond. Points $x$ and $y$ are given with $f(x) = 0$ and  $f(y) = 1$ and a particular path represents flipping variables in $S$ one at a time. Finally $z = x^{(S')}$ corresponds to picking some $z$ between $x$ and $y$; in this case, $f(z) = 1$, so $y$ would be updated to $z$.}
\label{fig:bs}
\end{figure}
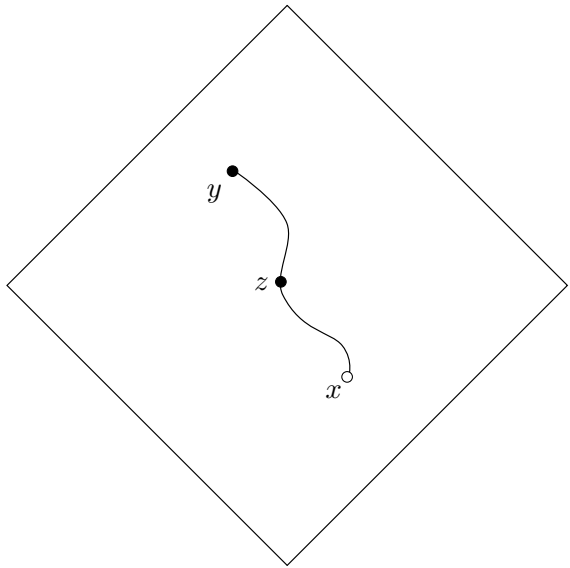

An edge violation occurs when two bi-chromatic edges \emph{collide}, i.e. they are in the same direction $i$ but one is monotone~and~the~other is anti-monotone. 
Thus, an algorithm may proceed by designing a subroutine for finding bi-chromatic edges and invoking this subroutine multiple times in hopes of finding a collision.
A subroutine for finding bi-chromatic edges that has been~widely~used~in~the~Boolean function property testing literature
  (e.g., \cite{B09, BB16,KS16})
  is \emph{binary search} (see Figure~\ref{fig:bs}):
\begin{enumerate}
\item Find two points $x, y \in \{0,1\}^n$
   with $f(x) \neq f(y)$, and let $S = \{ i \in [n] : x_i \neq y_i \}$. \vspace{-0.12 cm}
\item Pick a subset $S' \subset S$ of size $|S| / 2$,
  let $z = x^{(S')}$,\footnote{Here $x^{(S')} \in \{0, 1\}^n$ is the point obtained from $x$ by flipping its coordinates in $S'$;
  we also write $x^{(i)}$ for $x^{(\{i\})}$.} and query $f(z)$.\vspace{-0.12cm}
\item If $f(z) = f(x)$, let $x \leftarrow z$; if $f(z) = f(y)$, let $y \leftarrow z$. Repeat until
  $(x,y)$ is an edge.
\end{enumerate}
Clearly, the above procedure, if initiated with $f(x) \neq f(y)$,
  will always find a bi-chromatic edge~in some direction $i\in S$ with $O(\log n)$ queries.
One can further \emph{randomize} the subroutine by drawing $x$ and $y$ uniformly at random at the beginning
  and drawing $S'$ uniformly at random from $S$ in~each round.
Given an $f$, the binary search subroutine naturally induces a distribution
  over bi-chromatic edges of $f$.
A high-level question is: Can we analyze
  this distribution for functions $f$ that are $\eps$-far from unate?
   Can this strategy give better algorithms for finding an edge violation?

\begin{figure}[t!]
\centering
\begin{picture}(275,275)
    \put(0,0){\includegraphics[width=0.6\linewidth]{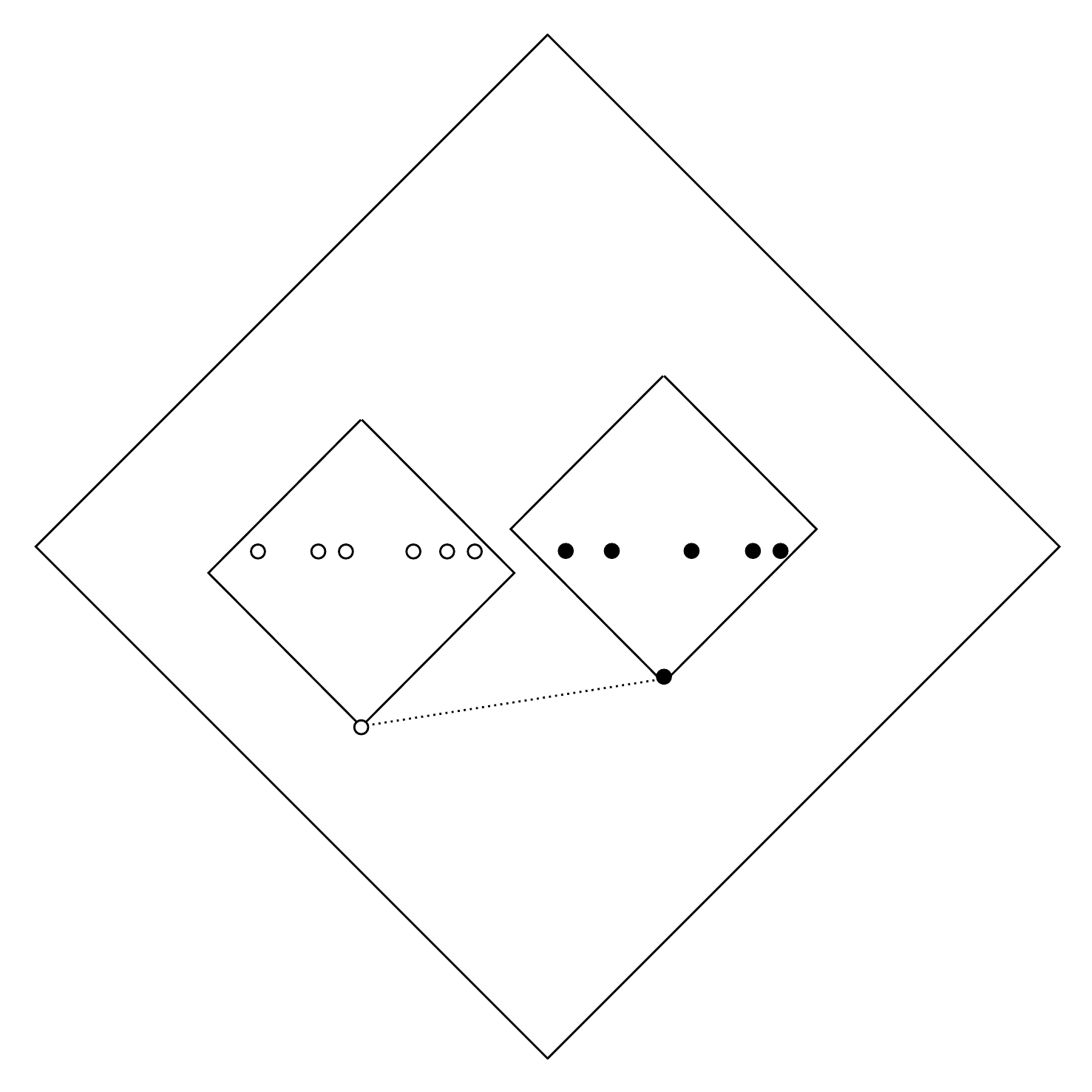}}
    \put(90, 85){$x$}
    \put(170, 98){$x^{(i)}$}
    \put(167, 187){$x^{(S)}$}
    \put(164, 145){$x^{(T_\ell)}$'s}
    \put(84, 127){$x^{(T_\ell)}$'s}
    \put(88, 175){$x^{(S\setminus \{i\})}$}
  \end{picture}
\caption{Pictorial representation of the adaptive edge search strategy, $\ESearch(x,S)$, for~finding bi-chromatic edges.
We consider the case when $(x,x^{(i)})$ is a bi-chromatic edge in direction~$i$~with $f(x)=0$ and $i\in S$.
The two sub-cubes in the picture above correspond to points that agree with $x$ outside of $S\setminus \{i\}$, and points that agree with
  $x^{(i)}$ outside of $S\setminus \{i\}$, respectively.
The points~$x^{(T_\ell)}$ sampled in $\ESearch(x,S)$
  lie in one of the sub-cubes according to whether $i \in T_{\ell}$ or not. 
Under certain conditions one can show that with high probability, all $T_\ell$'s with $f(x^{(T_\ell)})=1$~lie in~the~right sub-cube and furthermore, their intersection is exactly $\{ i \}$.
In this case, $\ESearch$ returns $i$.}
\label{fig:adaptive-esearch}
\end{figure}
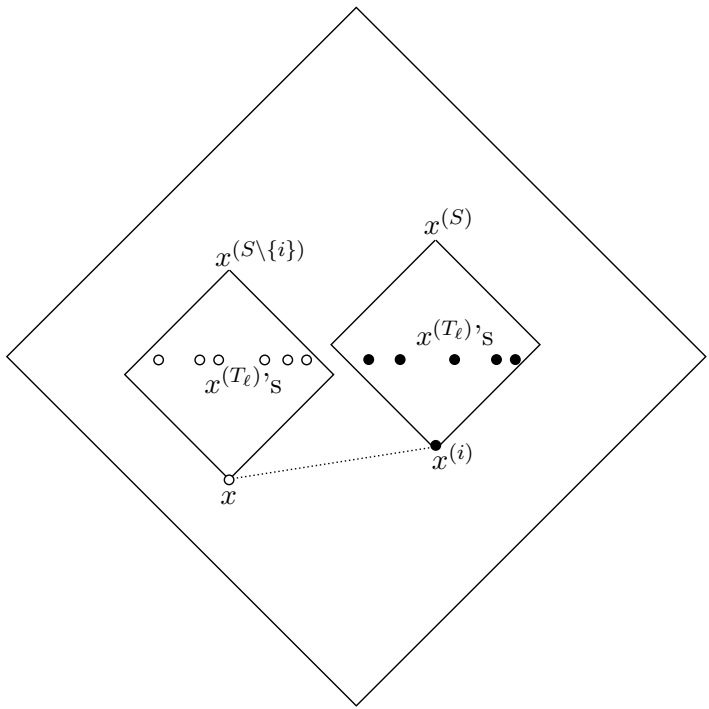

While we do not analyze the specific binary search strategy above in this paper, we introduce~a new kind~of edge search strategy, which we call \emph{adaptive edge search} and denote by $\ESearch$.~It is a crucial component~of our algorithm and allows for a relatively straightforward analysis. It~takes two inputs, a point $x\in \{0,1\}^n$
  and a nonempty set $S\subseteq [n]$,\hspace{0.01cm}\footnote{It is not important for the moment but later we will always   choose the size of $|S|$ to be smaller than $\sqrt{n}$.} with the goal of finding a bi-chromatic edge $(x,x^{(i)})$ for some $i\in S$
    (using $O(\log n)$ queries only).
The subroutine proceeds as follows:
\begin{enumerate}
\item Sample $L = O(\log n)$ subsets $T_1, \dots, T_L \subset S$ of size $|S| / 2$ uniformly, and query $f(x^{(T_\ell)})$.\vspace{-0.1cm}
\item Consider all $T_\ell$'s with $f(x^{(T_\ell)}) \neq f(x)$. If the intersection of such $T_\ell$'s consists of exactly \\one index $i\in S$, query $f(x^{(i)})$ and output $i$ if $f(x^{(i)})\ne f(x)$ (meaning that a bi-chromatic\\ edge
  $(x,x^{(i)})$ along direction $i$ has been found); otherwise return ``fail.''\end{enumerate}
See Figure \ref{fig:adaptive-esearch} for a pictorial representation.
While $\ESearch$ does not always finds~a bi-chromatic edge (unlike the binary search),
  its behavior is much easier to analyze.
Informally when $(x,x^{(i)})$ is a bi-chromatic edge  and $i\in S$ (otherwise it can never return $i$),
  $\ESearch(x,S)$ returns $i$ with high probability if
  (1)
  \emph{most} subsets $T\subset S$ of size $|S|/2$ with $i\notin T$
  have $f(x^{(T)})=f(x)$ and (2) \emph{most} $T\subset S$ of size~$|S|/2$ with $i\in T$
  have $f(x^{(T)})=f(x^{(i)})\ne f(x)$.
 (See Figure \ref{fig:adaptive-esearch} for an illustration.)


With the adaptive edge search in hand,
  the proof of Theorem \ref{thm:main}  proceeds in two steps.
For~the first step, we show that when $f$ is far from unate,  there must be
  ``many'' bi-chromatic edges $(x,x^{(i)})$ such that running $\ESearch$
  on $x$ paired with a random subset $S\subset [n]$ \emph{containing} $i$
  would lead to the discovery of  $(x,x^{(i)})$ with high probability.
There~are a lot of technical details hidden in the word ``many'': (i) subsets $S$ of different size
  contribute differently (intuitively, the larger~$S$, it is more likely for $S$ to
  contain $i$ when  $S$ is drawn from $[n]$ uniformly at random);
  (ii) we need~to balance the contribution from monotone and anti-monotone edges
  in the same direction  by taking their minimum. Intuitively, it will not help us find an edge violation
  if $\ESearch$ works well over many bi-chromatic edges in a direction $i$, but all these edges turn out to be
   monotone.~Following~the high-level discussion above,
  we formally introduce the notion of $\score^+_i$ and $\score^-_i $ for~a~Boolean
  function in Section \ref{sec:scores} (to measure the performance of $\ESearch$), and prove in Section \ref{sec:main2} that
\begin{align}\label{eqeqeq}
\sum_{i\in [n]} \min\big\{\score_i^+,\score_i^-\big\}=\widetilde{\Omega}(\eps^2),
\end{align}
when $f$ is $\eps$-far from unate. The proof of (\ref{eqeqeq}) heavily relies on the
   directed isoperimetric inequality~of \cite{KMS15} and its combinatorial implications for functions far
   from monotone
   (see Theorem \ref{thm:kms}).

In the second step, we present an algorithm that keeps calling the adaptive edge search (strategically),
  and show that it finds an edge violation with high probability, given (\ref{eqeqeq}).
At a high level,~it starts by sampling a set $S\subset [n]$ of certain size and a sequence of
  $K_+$ points $\{x_i\}$ from $\{0,1\}^n$.~Then it runs $\ESearch(x_i,S)$ for each $x_i$ and keeps the directions of monotone edges found in a set~$A$.
Next it samples $M$ subsets $T_i\subseteq S$ of certain size and for each $T_i$,  it
  samples $K_-$ points $\{y_{i,j}\}$ to run $\ESearch(y_{i,j},T_i)$. Similarly, it keeps the directions of anti-monotone edges found in $B$.
Finally, it outputs ``non-unate'' if $A\cap B\ne \emptyset$, i.e., an edge violation is found; otherwise, it outputs ``unate''.

The tricky part is the choices of the size of sets $S$ and $T$ and the three  parameters
  $M,K_+$ and $K_-$.
For technical reasons, our algorithm is split into two cases, depending on how the $\widetilde{\Omega}(\eps^2)$
  in (\ref{eqeqeq}) is achieved, e.g., what scale of $\min\{\score_i^-,\score_i^+\}$
  contributes the most in the sum.~The parameters are chosen differently in the cases and their proofs use slightly different
  techniques. \vspace{-0.25cm}

\paragraph{Organization.} \hspace{-0.2cm}We formally introduce the adaptive edge search subroutine in Section~\ref{sec:aes}. Next,~we introduce the notion of scores and state (\ref{eqeqeq}) in Lemma~\ref{main2} in Section~\ref{sec:scores}. We present the algorithm and its analysis in
  Section~\ref{sec:algorithm}, assuming Lemma~\ref{main2}. Finally we prove Lemma~\ref{main2} in Section~\ref{sec:main2}.

\begin{flushleft}
\setlength{\parindent}{15pt}
\end{comment}




\section{Preliminaries}\label{sec:preliminaries}

We use bold font letters such as $\bT$ and $\bx$ for random variables. Given $n\ge 1$, we write $[n]$ 
to denote $\{1, \dots, n\}$. Given a point $x$ in the Boolean hypercube $\{0, 1\}^n$ and $S \subset [n]$, 
  we use $x^{(S)}$ 
to denote the string obtained from $x$ by flipping each entry $x_i$ with $i \in S$. When $S = \{i \}$ is a
singleton, we write $x^{(i)}$ instead of $x^{(\{i\})}$ for convenience. Given $x, y \in \{0, 1\}^n$, $x \oplus y \in \{0, 1\}^n$ is their bit-wise~XOR.  


We define the \emph{distance} between two Boolean functions $f,g\colon\{0,1\}^n\rightarrow\{0,1\}$ using the~uniform
  distribution: $\dist(f,g):=\Pr_{\bx \sim\{0,1\}^n} [f(\bx)\ne g(\bx)]$.
The distance of $f$ to unateness is defined as~the minimum value of $\dist(f,g)$ over all unate functions $g$; 
  we say $f$ is \emph{$\eps$-far from unate} if its distance to unateness is at least $\eps$, or equivalently,
  $\dist(f,g)\ge \eps$ for all unate functions $g$.
  
We say an algorithm tests the unateness of Boolean functions if, given  $\eps$ and  query access to~a Boolean function $f$,
  (1) it ouputs ``unate'' with probability at least $2/3$ when $f$ is unate; and (2) it outputs ``non-unate''
  with probability at least $2/3$ when $f$ is $\eps$-far from unate.
We say the algorithm is one-sided if it always outputs ``unate'' when $f$ is unate. 

Recall that an \emph{edge violation} of unateness for $f$
  consists of a pair of bi-chromatic edges along the same direction, one being monotone and one being anti-monotone.
We remark that all algorithms discussed in this paper 
  output ``non-unate'' only when an edge violation is found among the queries they made.
We commonly refer to edge violations simply as violations. 




The \emph{total influence} $I_f$ of a Boolean function $f\colon\{0,1\}^n\rightarrow \{0,1\}$
  is the number of bi-chromatic edges of $f$ divided by $2^n$. 
We combine a lemma from \cite{KMS15} and 
  a unateness testing algorithm~of \cite{BCPRS17} to find an edge violation in a function  of \emph{high} total influence using $\widetilde{O}(\sqrt{n})$ queries  only.\footnote{Here we do not need
  to assume that $f$ is far from unate. Actually Lemma \ref{blabla1} implies that any $f$ with
  $I_f\ge 6\sqrt{n}$ cannot be unate, and the algorithm stated in Lemma \ref{hehelem} finds an edge violation with high probability.}
The proof can be found in Appendix \ref{sec:high-inf}.

\begin{lemma}\label{hehelem}
There is an $\widetilde{O}(\sqrt{n})$-query, non-adaptive 
  algorithm that, given any 
  $f \colon \{0, 1\}^n \to \{0, 1\}$ 
  with $I_f > 6\sqrt{n}$,
  finds an edge violation of $f$ to unateness with probability at least $2/3$.
\end{lemma}

Given Lemma \ref{hehelem}, it suffices for us to give an $\widetilde{O}(n^{3/4}/\eps^2)$-query algorithm that can
  find an edge violation of any function that is $\eps$-far from unate and satisfies $I_f\le 6\sqrt{n}$.


\section{Adaptive edge search}\label{sec:aes}

In this section, we
  introduce a new subroutine called \emph{adaptive edge search} ($\ESearch$) 
  which will be heavily used in our main algorithm for testing unateness.
We present the subroutine in Figure~\ref{fig:edge-search}.
It has query access to a Boolean function $f\colon \{0, 1\}^n \to \{0, 1\}$ and takes two inputs:
 $x \in \{0, 1\}^n$ is a point in the hypercube and
  $S \subseteq [n]$ is a nonempty set of even size. 

The goal of $\ESearch(x,S)$ is to find an index $i \in S$ such that $(x, x^{(i)})$ is a bi-chromatic edge in $f$.
It returns an index $i\in S$ if it finds one (note that $\ESearch$ always checks and makes sure that
  $(x,x^{(i)})$ is bi-chromatic before it outputs $i$), or returns ``fail'' if it fails to find one
  (which does not necessarily mean that none of the edges $(x,x^{(i)})$, $i\in S$, are bi-chromatic).
While a naive search would consider each $i \in S$ and query each $f(x^{(i)})$, as well as $f(x)$, incurring a cost of $|S|+1$~queries that can be expensive when $S$ is large,
  $\ESearch(x,S)$ only uses $L+2=O(\log n)$ queries, where we set the parameter $L=\lceil 4\log n\rceil$
  in Figure \ref{fig:edge-search}.

We will analyze the performance of $\ESearch$ in Section \ref{sec:main2}, where we show
  that, informally, under the assumption that $f$ is far from unate, $\ESearch$ succeeds in finding
  a bi-chromatic edge $(x,x^{(i)})$ for some $i\in S$
  for ``many'' input pairs $(x,S)$ with high probability. 
For now, we just record the following simple observation that follows from
  the description of $\ESearch$ in Figure~\ref{fig:edge-search}.
\begin{figure}[t!]
\begin{framed}
\noindent Subroutine \ESearch$(x, S)$
\vspace{-0.16cm}
\begin{flushleft}\noindent {\bf Input:} Query access to $f \colon \{0, 1\}^n \to \{0, 1\}$, $x \in \{0, 1\}^n$, 
  a nonempty set $S \subseteq [n]$ of even size.

\noindent {\bf Output:}  Either an index $i \in S$ with $f(x^{(i)}) \neq f(x)$, or ``fail.''

\begin{enumerate}

\item \vskip -.02in Query $f(x)$ and set $b\leftarrow f(x)$.\vspace{-0.06cm}
\item \vskip -0.02in If $|S| = 2$, pick one coordinate $\bi \in S$ uniformly at random. Query $f(x^{(\bi)})$ and\\ return 
$\bi$ if
$f(x^{(\bi)}) \neq b$; otherwise return ``fail.''\vspace{-0.06cm}

\item \vskip -0.02in  Sample $L=\lceil 4\log n\rceil$ subsets $\bT_1, \dots, \bT_L \subset S$ of size ${|S|}/{2}$ uniformly at random.
	 \\Query $f(x^{(\bT_\ell)})$ and set the output to be $\bb_\ell$ for each $\ell\in [L]$.
	 Let $\bC \subset S$ where\vspace{-0.06cm}
	\[ \bC = \bigcap_{\ell \in [L] \colon\hspace{-0.02cm}\bb_\ell \neq b} \bT_\ell\ \ \ \ \ \ \text{($\bC=\emptyset$ by default if $\bb_\ell=b$ for all $\ell$)}. \vspace{-0.06cm}\]
\item[] \vskip -0.06in If $\bC = \{ \bi \}$ for some $\bi$, query $f(x^{(\bi)})$ and return $\bi$ if $f(x^{(\bi)}) \neq b$;
otherwise return ``fail.''
\end{enumerate}
\end{flushleft}\vskip -0.14in
\end{framed}\vspace{-0.2cm}
\caption{Description of the adaptive edge search subroutine.\vspace{-0.15cm}} \label{fig:edge-search}
\end{figure}
\begin{fact}
$\ESearch(x,S)$ makes $O(\log n)$ queries and returns either an index $i$ or ``fail.''
Whenever it returns an index $i$, we have $i\in S$
  and $(x,x^{(i)})$ is a bi-chromatic edge in $f$.
\end{fact}

\section{Scores}\label{sec:scores}

In this section, we use the $\ESearch$ subroutine to introduce the notion of \emph{scores} for monotone and anti-monotone edges. 
We start with some notation. 

We consider some fixed function $f \colon \{0, 1\}^n \to \{0, 1\}$. For each $i \in [n]$, we let
$E_i^+$ denote the set of monotone edges in direction $i$ and 
$E_i^-$ denote the set of anti-monotone edges in direction $i$.
Let
\[ \Lambda = \left\lfloor \log_{2} \left(\frac{\sqrt{n}}{\log n} \right) \right\rfloor=\Theta(\log n) \]
be a parameter which will be used in the rest of the paper.
Given $i\in [n]$ and $j\in [ \Lambda]$, we let
  \[ \calP_{i, j} = \left\{ S \subset [n] \setminus \{i\} : |S| = 2^{j} - 1 \right\}. \]

We need the following definitions:

\begin{definition}[Good pairs]\label{def:goodpairs}
Let $(x,x^{(i)})$ be a monotone edge in $E_i^+$ for some $i\in [n]$ and let $S$ be a
  set in $\calP_{i,j}$ for some $j\in [\Lambda]$.
We say $(x,S)$ is a \emph{good pair for} $E_i^+$ if $\ESearch(x,S\cup \{i\})$~returns $i$
  with probability at least $1/2$ (i.e., running the adaptive edge search subroutine over $x$ and $S\cup \{i\}$ would
  help us discover the monotone edge $(x,x^{(i)})$ in $E_i^+$ with probability at least $1/2$).
\end{definition}

By definition, $(x,S)$ can be a good pair for $E_i^+$ only if $(x,x^{(i)})$ is a monotone edge. Additionally, if $(x, x^{(i)})$ is a monotone edge, then $(x, S)$ is always a good pair for all $S\in \calP_{i,1}$. This simply follows from the fact that, as $|S\cup\{i\}|=2$, $\ESearch(x,S\cup\{i\})$ will pick $i$ with probability $ {1}/{2}$ on line 2 and find the monotone edge $(x, x^{(i)})$. Next we use good pairs to define \emph{strong} points.

\begin{definition}[Strong points]
A point $x \in \{0, 1\}^n$ with $(x,x^{(i)})\in E_i^+$ is said to be $j$-\emph{strong} \emph{(}or~a $j$-\emph{strong point}\emph{)} for $E_i^+$,
  for some $j\in [\Lambda]$,
  if $(x,S)$ is a good pair for $E_i^+$ for at least $3/4$ of $S\in \calP_{i,j}$.
\end{definition}

Consider an $x \in \{0, 1\}^n$ that is a $j$-strong for $E_i^+$. Then, if we sample an $\bS$ from $\calP_{i,j}$ uniformly and
  run $\ESearch (x,\bS\cup \{i\})$, we will discover $(x,x^{(i)})\in E_i^+$ with probability $(3/4)\cdot (1/2)=3/8$.
Note that if $(x, x^{(i)})$ is a monotone edge, then $x$ is always $1$-strong. We also extend both definitions of \emph{good pairs} and $j$-\emph{strong} points to $\smash{E_i^-}$, so we may consider a good pair $(x, S)$ for $\smash{E_i^-}$, as well as,~a point $x\in \{0,1\}^n$ which is $j$-strong for $\smash{E_i^-}$,
  when $(x,x^{(i)})\in E_i^-$ is an anti-monotone edge.

For each $i\in [n]$ and $j\in [\Lambda]$, let $\score_{i,j}^+$ be
  the fraction of points that are $j$-strong for $E_i^+$:
$$
\score_{i,j}^+=\frac{ \text{the number of $j$-strong points for $E_i^+$}}{2^n}\in [0,1].
$$
Intuitively, when $\smash{\score_{i,j}^+}$ is high, it becomes easy to discover
  a monotone edge in direction $i$~using $\ESearch$ with $2^j$-sized sets  (as
  $|S\cup \{i\}|=2^j$ in Definition \ref{def:goodpairs}) without using too many queries.
This intuition will be made formal in the analysis of our main algorithm later in Section \ref{sec:main2}.

Finally we define $\score_i^+$ for each $i\in [n]$ as (recall that $2^j\le \sqrt{n}/\log n$ by the choice of $\Lambda$)
\begin{equation}\label{eheh}
\score_i^+=\max_{j\in [\Lambda]} \left\{\score_{i,j}^+\cdot  \frac{2^j}{\sqrt{n}}\right\}\in [0,1].
\end{equation}
Note that $\smash{\score_{i,j}^+}$'s are adjusted in (\ref{eheh}) with weights $2^j/\sqrt{n}$ before taking the maximum. 
Roughly speaking, this is done here to reflect the fact that with the same $\smash{\score_{i,j}^+}$, the larger $j$ is, the~easier it becomes to
  discover an edge in $E_i^+$ using sets of size $2^j$ in $\ESearch$. Consider some point $x \in \{0, 1\}^n$ which is $j$-strong for $E_i^+$. Then as noted above, if the algorithm samples a set $\bS \sim \calP_{i, j}$ uniformly at random  and runs $\ESearch(x, \bS \cup \{i \})$, the algorithm will observe $(x, x^{(i)})$ with probability $\frac{3}{8}$; however, the algorithm does not know $\calP_{i,j}$ or $i$. From the algorithm's perspective, there is a point $x \in \{0, 1\}^n$ with some bi-chromatic edge $(x, x^{(i)})$. The algorithm runs $\ESearch(x, \bS')$ for some set $\bS'$, and must hope that the set $\bS'$ of size $2^j$ contains $i$ and $\bS' \setminus \{ i \} \in \calP_{i,j}$ is a good pair. As $j$ increases, it becomes easier for $\bS'$ to include $i$.
(Again this will be made more formal in Section~\ref{sec:main2}).
We also extend  $\smash{\score_{i,j}^+}$, $\smash{\score_i^+}$ to
  $\smash{\score_{i,j}^-}$, $\smash{\score_i^-}$ for $E_i^-$.\vspace{-0.1cm}

\subsection{Plan for the proof of Theorem \ref{thm:main}\vspace{-0.06cm}}

The plan 
  is the following.
Let $f\colon\{0,1\}^n\rightarrow \{0,1\}$ be a function that is $\eps$-far from unate.
Our~goal is to give an $\widetilde{O}(n^{3/4}/\eps^2)$-query algorithm which
 finds an edge violation of $f$ with probability at least $2/3$.
By Lemma \ref{hehelem}, we may assume without loss of generality that $f$ also  satisfies $I_f \leq 6 \sqrt{n}$.
  
We rely on the following technical lemma for the scores of $f$ but delay its proof to Section \ref{sec:main2}.
\begin{lemma}\label{main2}
If $f\colon\{0,1\}^n\rightarrow \{0,1\}$ is $\eps$-far from unate and satisfies $I_f \leq 6\sqrt{n}$,
  then we have
\begin{equation}\label{maineq}
\sum_{i=1}^n \min\Big\{ \score_i^+,\score_i^-\Big\} \geq \Omega\left(\dfrac{\eps^2}{\log^8 n}\right).
\end{equation}
\end{lemma}

We  
  present our main $\widetilde{O}(n^{3/4}/\eps^2)$-query (adaptive) algorithm in the next section and show that, given any 
  function $f$ that satisfies~(\ref{maineq}),
  it finds an edge violation of $f$ with probability at least $2/3$.\vspace{-0.08cm}

\begin{flushleft}
\setlength{\parindent}{15pt}

\end{comment}

\section{Main algorithm and its analysis\vspace{-0.08cm}}\label{sec:algorithm}

We present our main algorithm that, given any Boolean function $f$ that 
  satisfies (\ref{maineq}), uses 
\begin{equation}\label{hehewiwi}
O\left(\frac{n^{3/4}}{\eps^2} \cdot \log^{16} n \cdot \log^{2}(n/\eps) \right) = \widetilde{O}(n^{3/4}/\eps^2)
\end{equation} 
queries 
  to find an edge violation of $f$ with probability at least $2/3$.

\subsection{Preparation: Bucketing scores}
\label{sec:bucketing}

We start with some preparation for the algorithm.
First, we use standard bucketing techniques to 
  make (\ref{maineq}) easier to use (while only losing a polylogarithmic factor in the sum).
Recall that
\[ \score_i^+ = \max_{j \in [\Lambda]} \left\{ \score_{i,j}^+ \cdot \dfrac{2^j}{\sqrt{n}} \right\} \quad 
\text{and} \quad \score_i^- = \max_{j \in [\Lambda]} \left\{ \score_{i,j}^- \cdot \dfrac{2^j}{\sqrt{n}} \right\}. \]
We will say that the $i$th direction is of \emph{type}-$(t,r)$, for some $t,r\in [\Lambda]$, if
\[ \score_i^+ = \score_{i,t}^+ \cdot \dfrac{2^t}{\sqrt{n}} \quad \text{and} \quad \score_i^- = \score_{i,r}^- \cdot \dfrac{2^r}{\sqrt{n}}. \]
As  $\Lambda=O(\log n)$,
  there are $O(\log^2 n)$ types.
From (\ref{maineq}) we know there is a pair $(t,r)$ such that
\begin{equation}
\label{eq:type-scores}
 \sum_{i:\hspace{0.05cm} \text{type-$(t,r)$}} \min\Big\{\score_i^+,\score_i^-\Big\}={\Omega} \left(\dfrac{\eps^2}{\log^{10}n}\right).
 \end{equation}
In the remainder of the section, 
  we fix such a type $(t,r)$ that satisfies (\ref{eq:type-scores}).
(Looking ahead, we~may assume that our algorithm knows $(t,r)$ as it can
  afford to try all $O(\log^2 n)$ possible pairs of $(t,r)$.)
  

Let $I^*\subseteq [n]$ be the set of all type-$(t,r)$ directions. We next divide $I^*$ into $\lceil 2\log(n/\eps)\rceil$ buckets according to $\min\{\score_{i}^+,\score_{i}^-\}$. An $i \in I^*$ lies in the $k$-th bucket if it satisfies
\[ \frac{1}{2^k} \leq \min\Big\{ \score_{i}^+, \score_{i}^-\Big\} \leq \frac{1}{2^{k-1}}. \]
 Note that some $i\in I^*$ may not lie in any bucket when $\min\left\{ \score_{i}^+, \score_{i}^-\right\} \leq  {\eps^2}/{n^2}$; however, all such $i\in I^*$ in total contribute at most $O(\eps^2/n)$ to the LHS of (\ref{eq:type-scores}), which is negligible compared to its RHS.
Since $k$ has $\lceil 2\log (n/\eps)\rceil=O(\log(n/\eps))$ possibilities, there exists an $h$ such that
\begin{equation}
\label{newmaineq}
 \sum_{i\in I^* \colon \hspace{-0.04cm}\text{bucket $h$}} \min \Big\{ \score_{i}^+, \score_{i}^-\Big\} \geq \Omega\left(\dfrac{\eps^2}{\log^{10} n \cdot \log(n/\eps)} \right).
 \end{equation}
Similarly we fix such an $h$ in the rest of the section (and assume later that the algorithm knows~$h$).
We also let $I\subseteq I^*$ be the indices of $I^*$ in bucket $h$. To simplify the notation, we let
  $H = 2^h$ and
\newcommand{\wteps}{\widetilde{\eps}}
\[ \wteps^2 = \dfrac{c\eps^2}{\log^{10} n\cdot  \log(n/\eps)}, \]
where we use $\wteps$ to hide the polylogarithmic factor in $\eps$ and $n$, and $c$ is some constant which ensures
\[ \sum_{i\in I} \min \Big\{\score_i^+, \score_i^-\Big\} \geq \wteps^2. \]
Given that $H=2^h$, we have $${1}/{H} \leq \min\Big\{ \score_{i}^+, \score_{i}^-\Big\} \leq  {2}/{H},\quad\text{for each $i\in I$,}$$ and 
  $|I|\cdot (2/H)\ge \wteps^2$
from (\ref{newmaineq}). This implies $H\leq 2|I|/\wteps^2=O(n/\wteps^2)$ 
  as $|I|\leq n$.
Moreover, using $$1/H\le \score_i^+ = \score_{i,t}^+\cdot (2^t/\sqrt{n})\le 2^t/\sqrt{n},$$
  we have $H 2^t\ge \sqrt{n}$ and similarly, $H 2^r\ge \sqrt{n}$.
 
 We summarize the above discussion with the following lemma.

\begin{lemma}
\label{lem:score-bucket}
\hspace{0.03cm}Suppose that $f$ satisfies \emph{(\ref{maineq})}.
\hspace{-0.05cm}Then there exist $t,r\in [\Lambda]$, $H=O(n/\wteps^2)$ as a power~of~$2$ with $H 2^t,H 2^r\ge \sqrt{n}$, and a nonempty $I \subseteq [n]$ of size $|I|\ge  H \wteps^2/2$
  such that every $i \in I$ satisfies
\[ \min\Big\{ \score_i^+, \score_i^- \Big\} = \min\left\{ \score_{i, t}^+ \cdot \dfrac{2^t}{\sqrt{n}},\hspace{0.06cm} \score_{i, r}^- \cdot \dfrac{2^r}{\sqrt{n}} \right\} \in \left[ {1}\big/{H}, {2}\big/{H}\right].\]
\end{lemma}


\subsection{Preparation: Informative sets}

\def\strong{\textsc{Strong}}

We introduce more notation and
  state Lemma~\ref{informative-lemma} that will be heavily used in the analysis of the algorithm. We will defer the 
  proof of Lemma~\ref{informative-lemma} to Subsection~\ref{proof-informative}.
Below $t,r$ and $H$ are considered as fixed parameters, and $I$ is a set of indices that satisfies
  the condition of Lemma \ref{lem:score-bucket}. 
We further assume that $t\ge r$; all our discussion below holds when $t<r$ by switching the roles of $t$ and $r$
  (and $E_i^+$ and $E_i^-$).
We start with some useful notation related to good pairs.

Recall  $(x,S)$ 
  is a good pair for $E_i^+$ (or $E_i^-$) if $(x,x^{(i)})$ is a monotone edge (or an~anti-monotone edge)
  and $\ESearch(x,S\cup \{i\})$ returns $i$ with probability at least $1/2$.
Given an $S \in \calP_{i, j}$, let 
\begin{align*}
&\good^+_i(S) = \big\{ x \in \{0, 1\}^n : (x, S) \text{ is a good pair for $E_i^+$} \big\},\quad\text{and}\\[0.4ex]
&\good^-_i(S)=\big\{ x \in \{0, 1\}^n : (x, S) \text{ is a good pair for $E_i^-$} \big\}.
\end{align*}
We also use
$$\goodf^+_i(S) = \dfrac{|\good_i^+(S)|}{2^n}\quad\text{and}\quad \goodf^-_i(S)=\dfrac{|\good_i^-(S)|}{2^n}$$
to denote the fraction of points in $\good^+_i(S)$ and $\good^-_i(S)$, respectively.
  
Recall that $x\in \{0,1\}^n$ is $j$-strong for $E_i^+$ (or $E_i^-$) if 
  $(x,S)$ is a good pair for $E_i^+$ (or $E_i^-$) for at least $3/4$ of sets $S\in \calP_{i,j}$.
Given an $i\in I$, we use $\strong_i^+$ to denote the set of
  $t$-strong points for $E_i^+$ and $\strong_i^-$ to denote the set of
  $r$-strong points for $E_i^-$.
By Lemma~\ref{lem:score-bucket}, we have
\begin{equation}\label{hehe22}
\score_{i,t}^+=\frac{|\strong_i^+|}{2^n}\ge \frac{\sqrt{n}}{H\cdot 2^t}
\quad\text{and}\quad
\score_{i,r}^-=\frac{|\strong_i^-|}{2^n}\ge \frac{\sqrt{n}}{H \cdot 2^r}.
\end{equation}

We define the following two parameters, which will be very important for the algorithm:
  \[ \alpha = \dfrac{|I| \cdot 2^t}{n} \quad \text{ and } \quad \beta = \dfrac{|I| \cdot 2^r}{n}.\]
These parameters measure the expectation of $|I \cap \bS|$ and $|I \cap \bT|$, respectively, when $\bS$ is a random subset of $[n]$ of size $2^t$ and $\bT$ is a random subset of $[n]$ of size $2^r$.

Finally we introduce the notion of 
  \emph{informative sets}.

\begin{definition}[Informative Sets]
\label{def:info}
We say a set $S\in \calP_{i,t}$ for some $i\in I$ 
  is \emph{informative} for the $i$th coordinate if both of the following two conditions hold:
\begin{enumerate}
\item $\goodf^+_i(S)\ge 0.1\cdot {\wteps^2}/({\alpha \sqrt{n}})$; and\vspace{-0.06cm}
\item $\goodf^-_i(T)\ge 0.1 \cdot {\wteps^2}/{(\beta \sqrt{n})}$
  for at least $0.1$-fraction of $(2^{r}-1)$-sized subsets $T$ of $S$.\\
We refer to $T \cup \{i \}$ as an $i$-\emph{revealing} set
  when $T$ has $\goodf^-_i(T)\ge 0.1 \cdot {\wteps^2}/{(\beta \sqrt{n})}$.
\end{enumerate} 
Additionally, we say the set $S \cup \{ i \}$ is \emph{$i$-informative} if $S$ is informative for the $i$th coordinate.
\end{definition}

To gain some intuition, if the algorithm is given a set $S\in \calP_{i,t}$ for some $i\in I$ 
  that is informative for the $i$th coordinate, then it can use $S$ and $i$ to find a violation 
  along the $i$th direction  as follows. 
\begin{flushleft}\begin{enumerate}
\item Sample $O( \alpha \sqrt{n}/{\wteps^2} )$ points $\bx \in \{0, 1\}^n$ uniformly and 
  run $\ESearch(\bx, S \cup \{i\})$.\vspace{-0.12cm} 
\item Sample a subset $\bT  \subseteq S$ of size $ 2^r - 1 $, sample $O ( {\beta \sqrt{n}}/{\wteps^2} )$ points $\by \in \{0, 1\}^n$
  uniformly at random, and then run $\ESearch(\by, \bT \cup \{i \})$. 
\end{enumerate}\end{flushleft}
We find a violation if we find a monotone edge in direction $i$ in step 1 and an anti-monotone edge
  in direction $i$ in step 2.
By  Definition~\ref{def:info}, 
  this occurs with probability $\Omega(1)$.
Of course, the~algorithm does not have knowledge of $S$ and $i$, so we need to incorporate other ideas; however, the intuition is that informative sets can help reveal edge violations efficiently using the $\ESearch$ subroutine.

The key will be to show that there are many informative sets for each $i\in I$, which we do in the following lemma using standard averaging arguments,
  but delay its proof to Section \ref{proof-informative}.
\begin{lemma}\label{informative-lemma}
For each $i\in I$, at least $1/8$ of sets $S\in \calP_{i,t}$ are informative
  for the $i$th coordinate.
\end{lemma}

\subsection{Cases of the main algorithm}

We are now ready to describe the main algorithm (which is one-sided and returns ``non-unate'' \emph{only} when it finds an edge violation of unateness). 
As mentioned earlier, we focus on the case when $f$ satisfies (\ref{maineq})
  and show that for any such $f$, the algorithm finds an edge violation with probability at least $2/3$.
We assume that the algorithm knows the parameters $r,t$ and $H$ from Lemma \ref{lem:score-bucket} 
(algorithmically, we just try all possibilities for these parameters, which incurs a factor of $O(\log^2n\cdot  \log(n/\eps))$ in the final query complexity).
Let $I\subseteq [n]$ be the set promised in Lemma~\ref{lem:score-bucket} (note that~algorithm has~no knowledge about $I$).
We also assume that $t\ge r$; if not, one can switch the roles of monotone and anti-monotone edges by
  running the algorithm on $\smash{g(x)=f(x\oplus 1^n)}$, where $1^n$ is the all-1's string.



\def\algfirst{\texttt{Alg-Case-1}}

\subsection{Case 1: $\alpha \geq \log^2 n$}

In this case, we expect a random set $\bS$ of size $2^t$ to have intersection with (the unknown) $I$ of size at least $\log^2 n$. 
The algorithm, \algfirst, is presented in Figure \ref{fig:case1} with the following parameter:
\[ M = \left\lceil \dfrac{\sqrt{\alpha n}}{\wteps^2} \cdot \log^3 n \right\rceil.  \]

\begin{figure}[t!]
\begin{framed}
\noindent Subroutine $\algfirst$, handling the case when $\alpha \geq \log^2 n$
\vspace{-0.16cm}
\begin{flushleft}\noindent {\bf Input:} Query access to $f \colon \{0, 1\}^n \to \{0, 1\}$

\noindent {\bf Output:}  Either ``unate,'' or two edges constituting an edge violation for $f$.

\begin{itemize}

\item[] \vskip -.03in \hspace{-0.2cm}Repeat the following $O(1)$ times for some sufficiently 
  large constant: \vspace{-0.06cm}

\begin{enumerate}

\item \vskip -.04in Sample a set $\bS$ of size $2^{t}$ from $[n]$ uniformly at random.

\item  Repeat $M$ times:

\begin{itemize}
\item Sample an $\bx \in \{0, 1\}^n$ uniformly at random and run $\ESearch(\bx, \bS)$.
\end{itemize}

\item Let $\bA$ be the set of $i \in [n]$ such that a monotone edge in direction $i$ is found.

\item Repeat $M$ times:
\begin{itemize}
\item Sample a subset $\bT \subseteq \bS$ of size $2^r$\vspace{0.06cm} uniformly at random, as well as $\by \in \{0, 1\}^n$ uniformly and run $\ESearch(\by, \bT)$.\vspace{0.08cm}
\end{itemize}
\item Let $\bB$ be the set of $i \in [n]$ such that an anti-monotone edge in direction $i$ is found. 
\item If $\bA \cap \bB\ne \emptyset$, output an edge violation of $f$ to unateness. 
\end{enumerate}
\item[] \vskip -0.06in \hspace{-0.2cm}If we have not found any edge violation in line 6, output ``unate.'' 
\end{itemize}
\end{flushleft}\vskip -0.14in
\end{framed}
\caption{Description of the \algfirst~for Case 1 of the algorithm.} \label{fig:case1}
\end{figure}


\begin{fact}[Query complexity]
The number of queries used by $\emph{\algfirst}$~is \emph{(}using $\alpha\le \sqrt{n}$\emph{)}
$$
O(1) \cdot \left( M + M \right) \cdot O(\log n) = O\left( \dfrac{\sqrt{ \alpha n}\cdot \log^{4} n }{\wteps^2}\right)=O\left(\frac{n^{3/4}\cdot \log^{14} n \cdot \log (n/\eps)}{\eps^2}\right).
$$
\end{fact}

\paragraph{Correctness of Case 1:}
\hspace{-0.2cm}Below, we prove that \algfirst~finds a violation with high probability.
We split the proof into two lemmas. The first, Lemma~\ref{lem:1-1}, shows that a certain condition is satisfied with constant probability in each iteration of Step 1 of \algfirst.
The second, Lemma~\ref{lem:case1-1}, shows that if the condition of Lemma~\ref{lem:1-1} is satisfied at Step 1 of \algfirst, then the algorithm finds a violation with high probability.

\begin{lemma}\label{lem:1-1}
Let $\bS$ be a $2^t$-sized subset drawn from $[n]$ uniformly  at random and let $\bI_{\bS} \subseteq  I \cap \bS$~be the set of $i\in I\cap \bS$ such that $\bS$ is $i$-informative. Then ${\alpha}/{10} \leq |\bI_{\bS}| \leq 4\alpha$ with probability $\Omega(1)$. 
\end{lemma}

\begin{proof}
Recall that $\alpha$ is the expected size of $I \cap \bS$.~As a result of $\alpha \geq \log^2 n$, the fraction of $S\subset [n]$ of size $2^t$ with
  $|S\cap I|> 4\alpha $ is at most $\exp(-\Omega(\log^2 n))$ (see Lemma \ref{appendix1} in Appendix~\ref{apphehe} for a formal proof).
In the rest of the proof, we let 
\[ \calS = \big\{S \subset [n] : |S| = 2^t\ \text{and}\ |S \cap I| \le 4\alpha \big\}. \]

We define a  bipartite graph $H^*$: vertices on the two sides correspond to $I$ and $\calS$, respectively;
$(i,S)$ is an edge if $S$ is $i$-informative.
By Lemma \ref{informative-lemma}, the degree of each $i\in I$ is at least
$$
\frac{1}{8} \cdot |\calP_{i,t}|-\exp\left(-\Omega(\log^2 n)\right)\cdot
{n\choose 2^t}\geq \frac{1}{9}\cdot |\calP_{i,t}| = \frac{1}{9} \cdot {n-1\choose 2^t-1}.
$$
Let $\gamma$ denote the fraction of $S\in \calS$ (among $\calS$) with degree at least $\alpha/10$ in $H^*$.
On the one hand,~the number of edges in $H^*$ is at least (counting from the $I$-side
  and using $|\calS|\le {n\choose 2^t}$)
$$
|I|\cdot \frac{1}{9} \cdot {n-1\choose 2^t-1} \ge \frac{1}{9} \cdot |I|\cdot \frac{2^t}{n}\cdot |\calS|
=  \frac{1}{9} \cdot \alpha |\calS|.
$$
On the other hand, the number of edges in $H^*$ is at most (counting from the $\calS$-side)
$$
\gamma|\calS|\cdot 4\alpha+(1-\gamma)|\calS|\cdot (\alpha/10)=\alpha |\calS|\cdot \left(\frac{39\gamma}{40}+\frac{1}{10} \right).
$$
As a result, $\gamma=\Omega(1)$. Since $\calS$ consists of
  $(1-o(1))$-fraction of all sets $S\subset [n]$ of size $2^t$,
  the set $\bS$ sampled in Step 1 of \algfirst~lies in $\calS$ and has degree between $\alpha/10$ and $4\alpha$ with
  probability at least $\Omega(1)$.
\end{proof}

\begin{lemma}
\label{lem:case1-1}
Suppose \emph{\algfirst}~samples a set $S$, and let $I_S \subseteq I\cap S$ be the set of $i$ such that
  $S$ is $i$-informative. If ${\alpha}/{10} \leq |I_S| \leq 4 \alpha$, \emph{\algfirst}~finds an edge violation  with probability $1 - o(1)$.
\end{lemma}

The lemma is further divided into simple claims. We consider a fixed set $S \subset [n]$ of size $2^t$ such that $\alpha/10\le |I_S|\le 4\alpha$. We let \algfirst~run up to Step 3, and let $\blambda = |\bA \cap I_S|$.

\begin{claim}\label{cl:1-1}
After the $M$ iterations of Step 2 in \algfirst, $\blambda \geq \sqrt{\alpha}$ with probability $1-o(1)$.
\end{claim}

\begin{proof}
We divide the $M$ samples of Step 2 into $\sqrt{\alpha}$ batches, each of $$M / \sqrt{\alpha} = \Omega\left( \dfrac{\sqrt{n}}{\wteps^2}\cdot \log^{3} n\right)$$ rounds of Step 2. For batch $\ell$, we let $\bX_\ell$ denote the indicator random variable for the event that at the start of the $\ell$th batch, $|\bA \cap I_S| < \sqrt{\alpha}$, and the $\ell$th batch fails to discover a monotone edge along a 
  new direction in $I_{S}\setminus \bA$. We will prove that all $\bX_\ell$ are 0 with probability $1-o(1)$. The lemma follows.

Suppose that at the start of the $\ell$th batch, $|\bA \cap I_S| < \sqrt{\alpha}$. Then consider the auxiliary bipartite graph $H^*$: vertices on the left-hand side consist of all points $x \in \{0, 1\}^n$; vertices on the right-hand side consist of indices of $I_{S}$; an edge $(x, i)$ is present if $x$ and $S \setminus \{i \}$ forms a good pair for $E_i^+$. 

Note that every vertex on the right-hand side has degree at least $0.1 \cdot \wteps^2 / (\alpha \sqrt{n}) \cdot 2^n$; vertices on the left-hand side have degree at most $2$ (because $(x,i)$ is an edge only
  if \ESearch$(x,S)$ returns $(x,x^{(i)})$ with probability at least $1/2$). Thus, the fraction of points on the left-hand side which are connected to at least one vertex on right-hand side that is currently not in $\bA$ is at least 
\[ \Big(|I_S|-\sqrt{\alpha}\Big) \cdot \dfrac{0.1 \cdot \wteps^2}{\alpha \sqrt{n}} \cdot \dfrac{1}{2} 
\ge \frac{|I_S|}{2} \cdot \dfrac{0.1 \cdot \wteps^2}{\alpha \sqrt{n}} \cdot \dfrac{1}{2}
= \Omega\left( \dfrac{\wteps^2}{\sqrt{n}}\right).\]
Thus, by the setting of $M / \sqrt{\alpha}$, we discover a new index in $I_S$ during the 
  $i$th batch with probability  at least $1-\exp(-\log^3 n)$, and we may union bound over the $\sqrt{\alpha} \leq n^{1/4}$ batches.
\end{proof}

We now turn to proving that in Step 4 of \algfirst, we will discover  an anti-monotone edge which, together with a monotone edge from Step 2, forms an edge violation. We divide the proof into two cases, corresponding to the value of $\beta$. The first two claims correspond to the case when $\beta \geq \alpha\log^2 n / \lambda$, and the second two claims correspond to the case when $\beta < \alpha\log^2 n / \lambda$. 

\begin{claim}
\label{lem:case-1-2}
Suppose $\beta \geq \alpha\log^2 n  / \lambda$. Let $A$ be a fixed set after running up to Step 4 of \emph{\algfirst} satisfying $|A \cap I_S| =\lambda \geq \sqrt{\alpha}$. Then with probability at least $\Omega(1)$ over the draw of a $2^r$-sized random subset $\bT$~of~$S$, the 
  number of indices $i\in A \cap I_S$ such that $\bT$ is $i$-revealing is at least 
  $\beta/(100\sqrt{\alpha})$.
\end{claim}
\begin{proof}
First we let $\calT$ denote the following set:
\[ \calT = \big\{ T \subseteq S : |T| = 2^r\ \text{and}\ |T \cap A \cap I_S| \leq 4 \lambda \cdot \beta / \alpha \big\}. \] 
 The expectation of $|\bT \cap A \cap I_S|$ when $\bT$ is a random subset of $S$ of size $2^r$ is at most $\lambda \cdot \beta / \alpha$. 
Since $\lambda \cdot \beta / \alpha \geq \log^2 n$ (by assumption),  $\calT$ consists of all but an $ \exp(-\Omega(\log^2 n)) $-fraction of subsets~of $S$ of size $2^r$ 
(See Lemma \ref{appendix1} in Appendix~\ref{apphehe} for a formal proof). 

Next consider a bipartite graph  $H^*$: vertices on the LHS correspond to 
$i \in A \cap I_S$; vertices~on the RHS correspond to sets $T \in \calT$;
  $(i,T)$ is an edge if $T$ is $i$-revealing. Note that since $i \in A \cap I_{S}$, $S$
  is $i$-informative and thus, 
  the degree of each $i$ in $H^*$ is at least
  \[ 0.1 \cdot {2^t - 1 \choose 2^r - 1} - \exp\left( -\Omega(\log^2n) \right) \cdot {2^t \choose 2^r} \geq \frac{1}{20} \cdot { 2^t - 1\choose 2^r - 1}. \] 
We show below that many $T\in \calT$ have degree at least $\lambda \beta/(100\alpha)$.
To this end let $\gamma$ be the fraction of $T\in \calT$ on the RHS which have degree at least $\lambda \beta/(100 \alpha)$ (among all $2^r$-subsets of $S$).
Then 
\begin{align*}
{\gamma} \cdot {2^t\choose 2^r}\cdot \frac{4\lambda \beta}{ \alpha}
+{(1-\gamma)}\cdot {2^t\choose 2^r}
\cdot \frac{\lambda \beta}{100 \alpha} \ge \lambda \cdot \frac{1}{20} \cdot {2^t-1\choose 2^r-1},
\end{align*}
and because $2^t / 2^r = \alpha / \beta$, canceling the factors, we obtain
$\gamma = \Omega(1)$.
This means that $\Omega(1)$-fraction of $2^r$-subsets of $S$ 
  have degree at least ${\lambda\beta}/({100\alpha}) \geq \beta / (100 \sqrt{\alpha})$, and the claim follows.
\end{proof}

\begin{claim}
Suppose $\beta \geq \alpha\log^2 n  / \lambda$. Consider \emph{\algfirst}~run up to Step 4, and we let $|A \cap I_S| = \lambda \geq \sqrt{\alpha}$. After $M$ iterations of Step 4 in \emph{\algfirst}, 
  we have $A \cap \bB \neq \emptyset$ with probability $1-o(1)$.
\end{claim}

\begin{proof}
Note that with probability $\Omega(1)$, $|\bT \cap A \cap I_S| \geq \beta / (100 \sqrt{\alpha})$. Similar  to the proof of Claim~\ref{cl:1-1} we let $H^*$ denote an auxiliary bipartite graph: vertices on the left-hand side correspond to points $x \in \{0, 1\}^n$; vertices on the right-hand side correspond to indices $i\in \bT \cap A \cap I_S$; $(x, i)$ are connected if $(x, T \setminus \{i \})$ forms a good pair for $E_i^-$. Note that each $i$ on the right-hand side has degree~at~least $0.1 \cdot \wteps^2 / (\beta \sqrt{n}) \cdot 2^n$; each point $x$ on the left-hand side has degree at most 2. Hence the fraction of points on the left-hand size which are connected to points on the right-hand side is at least 
\[ \dfrac{\beta}{100 \sqrt{\alpha}} \cdot \dfrac{0.1 \cdot \wteps^2}{\beta \sqrt{n}} \cdot \frac{1}{2} = \Omega\left( \dfrac{\wteps^2}{\sqrt{\alpha n}}\right). \]
By our choice of $M$, we find a violation with probability at least $1-o(1)$.
\end{proof}

This finishes the case when $\beta \geq \alpha\log^2 n / \lambda$. Now we consider the case 
  when $\beta <\alpha \log^2 n / \lambda$. 

\begin{claim}
\label{lem:case-1-3}
Suppose $\beta< \alpha\log^2 n / \lambda$. Let $A$ be a fixed set after running up to Step 4 of \emph{\algfirst} with $|A \cap I_S| =\lambda \geq \sqrt{\alpha}$. Then with probability at least $\Omega(\beta / (\sqrt{\alpha} \log^2 n)$ over the draw of a $2^r$-sized random subset $\bT$~of~$S$, there is at least one index $i\in A \cap I_S$ such that $\bT$ is $i$-revealing.
\end{claim}
\begin{proof}
First we let $\calT$ denote the following set:
\[ \calT = \big\{ T \subset S : |T| = 2^r\ \text{and}\ |T \cap A \cap I_S| \leq 4\log^2 n \big\}. \] 
Since $|A \cap I_S| = \lambda$, the expectation of $|A \cap I_S \cap \bT|$ when $\bT$ is a random $2^r$-subset of $S$  is at most $\lambda\beta/\alpha<\log ^2n$. 
As a result, $\calT$ consists of all but an $ \exp(-\Omega(\log^2 n)) $-fraction of subsets~of $S$ of size $2^r$
(see Lemma \ref{appendix1} in Appendix~\ref{apphehe} for a formal proof). 

Consider a bipartite graph  $H^*$: vertices on the LHS correspond to 
  indices $i \in A \cap I_S$; vertices~on the RHS correspond to sets $T \in \calT$;
  $(i,T)$ is an edge if $T$ is $i$-revealing. Note that since $i \in A \cap I_{S}$, $S$ is 
   $i$-informative and thus,
  the degree of each $i$ in $H^*$ is at least
  \[ 0.1 \cdot {2^t - 1 \choose 2^r - 1} - \exp\left( -\Omega(\log^2n) \right) \cdot {2^t \choose 2^r} \geq \frac{1}{20} \cdot { 2^t - 1\choose 2^r - 1}. \] 
On the other hand, the degree of each vertex on the right-hand side is at most $4 \log^2 n$. Therefore, the fraction of vertices on the RHS (among all $2^r$-subsets of $S$) which are not isolated is at least
\[  {|A \cap I_S|}\cdot \frac{1}{20} \cdot \binom{2^t - 1}{2^r - 1} \cdot \frac{1}{4\log^2 n}\cdot \frac{1}{{2^t\choose 2^r}} \geq \dfrac{\lambda \cdot \beta}{20 \cdot \alpha \cdot 4\log^2 n} \geq \Omega\left( \dfrac{\beta}{\sqrt{\alpha} \cdot \log^2 n}\right),\]
where the last inequality used $\lambda\ge \sqrt{\alpha}$.
This finishes the proof of the claim.
\end{proof}

\begin{claim}
Suppose $\beta <\alpha\log^2 n  / \lambda$. Consider \emph{\algfirst}~run up to Step 4 and we let $|A \cap I_S| = \lambda \geq \sqrt{\alpha}$. After $M$ iterations of step 4 in \emph{\algfirst}, we have $A \cap \bB \neq \emptyset$ with probability $1-o(1)$.
\end{claim}

\begin{proof}
From Claim~\ref{lem:case-1-3}, with probability $\Omega ({\beta}/({\sqrt{\alpha} \log^2 n}))$,   there exists an index $i \in A \cap I_S$  such that $i\in \bT$ and $\bT$ is $i$-revealing. When such a set $T$ is sampled, since $T$ is $i$-revealing, there exist at least $0.1 \cdot \wteps^2 / (\beta \sqrt{n}) \cdot 2^n$ points $y$ for which $\ESearch(y, T)$ returns an anti-monotone edge in direction $i$ with probability at least $1/2$. Thus, with probability at least $$\Omega\left( \dfrac{\beta}{\sqrt{\alpha} \log^2 n} \cdot \dfrac{\wteps^2}{\beta \sqrt{n}} \right)=\Omega\left(
\dfrac{\wteps^2}{\sqrt{\alpha n}\cdot \log^2 n}\right)$$ over the draw of $\bT$, $\by$, and the randomness of $\ESearch$, we find a violation to unateness. This finishes the proof by our choice of the parameter $M$.
\end{proof}

\newcommand{\algsecond}{\texttt{Alg-Case-2}}
\subsection{Case 2: $\alpha < \log^2 n$}

In this case, we expect a random set $\bS$ of size $2^t$ and a random set $\bT$ of size $2^r$ to have a relatively small intersection with (the unknown) $I$. 
We can actually achieve an $\widetilde{O}(\sqrt{n} /\eps^2)$ query complexity~in this case. The algorithm, \algsecond, is presented in Figure \ref{fig:case2} with the following parameters:
\[ K =\left\lceil\dfrac{\log^3 n}{\alpha}\right \rceil=\Omega(\log n) \quad \text{and} \quad M = \left\lceil \dfrac{\alpha \sqrt{n} \cdot \log n}{\wteps^2} \right\rceil=\Omega(\log n), \]
using $\alpha< \log^2 n$ and 
  $\alpha=|I|2^t/n$, $|I|\ge H\wteps^2/2$ and $H2^t\ge \sqrt{n}$ from Lemma \ref{lem:score-bucket}.


\begin{figure}[t!]
\begin{framed}
\noindent Subroutine $\algsecond$, handling the case when $\log^2 n> \alpha \geq \beta$
\vspace{-0.16cm}
\begin{flushleft}\noindent {\bf Input:} Query access to $f \colon \{0, 1\}^n \to \{0, 1\}$

\noindent {\bf Output:}  Either ``unate,'' or two edges constituting an edge violation for $f$.

\begin{itemize}

\item[] \vskip -.02in \hspace{-0.2cm}Repeat the following $K$ times: \vspace{-0.08cm}

\begin{enumerate}

\item \vskip -.02in Sample a set $\bS$ of size $2^{t}$ from $[n]$ uniformly at random.

\item  Repeat $M$ times:

\begin{itemize}
\item Sample an $\bx \in \{0, 1\}^n$ uniformly at random and run $\ESearch(\bx, \bS)$.
\end{itemize}

\item Let $\bA$ be the set of $i \in [n]$ such that a monotone edge in direction $i$ is found.

\item Repeat $M$ times:
\begin{itemize}
\item Sample a subset $\bT \subseteq \bS$ of size $2^r$ \vspace{0.06cm}uniformly at random, as well as a point $\by \in \{0, 1\}^n$ uniformly and run $\ESearch(\by, \bT)$.\vspace{0.08cm}
\end{itemize}
\item Let $\bB$ be the set of $i \in [n]$ such that an anti-monotone edge in direction $i$ is found. 
\item If $\bA \cap \bB\ne \emptyset$, return a violation of $f$ to unateness found. 
\end{enumerate}
\item[] \vskip -0.045in \hspace{-0.2cm}If we have not found any violation in line 6, return ``unate.'' 
\end{itemize}
\end{flushleft}\vskip -0.14in
\end{framed}
\caption{Description of the \algsecond~for Case 2 of the algorithm.} \label{fig:case2}
\end{figure}


\begin{fact}[Query Complexity]
The number of queries used by $\emph{\algsecond}$~is
$$
K \cdot \left( M + M \right) \cdot O(\log n) = O\left( \dfrac{\sqrt{n} \cdot \log^{5} n }{\wteps^2}\right).
$$
\end{fact}

\noindent\textbf{Correctness of Case 2:}
Below we prove that \algsecond~finds a violation with high probability.
We divide the proof into two lemmas. The first lemma obtains a sufficient condition for finding~an edge violation for $f$, and the second shows that the condition is satisfied with high probability.

\begin{lemma}
\label{lem:case2-1}
Suppose \emph{\algsecond}~starts with a set $S$ that is $i$-informative for some $i \in I$. Then during this loop, it finds an edge violation for $f$ along the $i$th direction with probability $1 - o(1)$.
\end{lemma}

\begin{proof}
Let $S'=S\setminus \{i\}\in \calP_{i,t}$.
Since $S'$ is informative for the $i$th coordinate, we have
\begin{equation}\label{heha} \goodf_{i}^+(S') \geq 0.1 \cdot \dfrac{\wteps^2}{\alpha \sqrt{n}} \quad \text{ and } \quad \goodf_{i}^-(T') \geq 0.1 \cdot \dfrac{\wteps^2}{\beta \sqrt{n}} 
\end{equation}
for at least $0.1$-fraction of $(2^r-1)$-sized subsets $T' \subset S'$.
We show below that $i\in \bA \cap \bB$ at the end of the loop with probability at least $1-o(1)$. 

First, by the definition of good pairs, every time an $x$ sampled in Step 2 lies in $\good_i^+(S')$, we have 
  $\ESearch(x,S)$ outputs the monotone edge $(x, x^{(i)})$ with probability at least $1/2$.
Using our choice of $M$ 
  we have $i\in \bA$ at the end of Step 3 in this loop with probability at least $1 - o(1)$. 

Next,  the number of $(2^r-1)$-sized subsets $T'$ of $S'$ satisfying (\ref{heha}) is at least
$$
0.1\cdot {2^t-1 \choose 2^r-1}.
$$
As a result, $T'\cup\{i\}$ obtained from such $T'$ consist of at least an
$$
\Omega\left( {{2^t - 1\choose 2^r - 1}}\Big/{{2^t \choose 2^r}}\right)=
\Omega\left(\frac{2^r}{2^t}\right)=\Omega\left(\frac{\beta}{\alpha}\right)
$$
fraction of $2^r$-subsets of $S$. When such a $T'\cup \{i\}$ is sampled in Step 4, the fraction of points $y$ that can help us discover an anti-monotone edge in direction $i$ using $\ESearch(y,T'\cup \{i\})$ is at least $\Omega( {\wteps^2}/({\beta \sqrt{n}}))$. Thus we observe an anti-monotone edge in direction $i$ with probability $\Omega( {\wteps^2}/{\alpha \sqrt{n}})$ 
  over the draw of each pair of $\bT$ and $\by$ in Step 4. 
So by our choice of $M$, we observe such a violation with probability at least $1 - o(1)$.
This finishes the proof of the lemma. 
\end{proof}

\begin{lemma}
The probability of a random $2^t$-sized subset $\bS$ being
   $i$-informative for some $i\in I$ is  at least
$\Omega({\alpha}/{\log^2 n}).$
\end{lemma}
\begin{proof}
We lowerbound the number of $S\subset [n]$ of size $2^t$
  that are $i$-informative for some $i \in I$.

Using $\alpha < \log^2 n$, the fraction of $2^t$-subsets $S$ with $| S\cap I|\ge 4\log^{2} n$ is at most $\exp(-\Omega(\log^{2} n))$ (see Lemma \ref{appendix1} in Appendix~\ref{apphehe} for a formal proof).
Next we let 
\[ \calS =\big\{ S \subset [n] : |S| = 2^t\ \text{and}\ |S\cap I|\le 4\log^{2} n \big\}. \]
We consider the following auxiliary bipartite graph $H^*$:
vertices on the LHS  are $i\in I$; vertices on the
  RHS are $S\in \calS$; a pair $(i,S)$ is an edge if $S$ contains $i$ and is $i$-informative. 
Thus, it suffices to show that many $S \in \calS$ on the RHS of $H^*$ are not isolated.

By Lemma \ref{informative-lemma}, for each $i\in I$, at least $1/8$ of $S'\in\calP_{i, t}$ 
  are informative for the $i$th direction.
If $S' \in \calP_{i, t}$ is one such set then $(i, S' \cup \{ i\})$ is an edge 
  when $S'\cup\{i\}\in \calS$. 
So the degree of $i$ is at least
$$
\frac{1}{8} \cdot |\calP_{i,t}|-\exp\left(-\Omega(\log^{2} n)\right)\cdot {n \choose 2^t}=\Omega\big(|\calP_{i,t}|\big)=\Omega\left({n-1\choose 2^t-1}\right).
$$
On the other hand, each $S \in \calS$ has degree at most $4\log^{2} n$, since $|S \cap I|\leq 4\log^{2} n$
  for every $S\in \calS$. Thus, the number
  of vertices on the RHS that are not isolated is at least
$$
|I|\cdot \Omega\left({n-1\choose 2^t-1}\right)\cdot \frac{1}{4\log^{2} n}
\geq \Omega\left(\dfrac{ |I|}{ \log^{2} n} \cdot \dbinom{n-1}{2^t-1}\right).
$$
As a result, the probability of a random $2^t$-sized set $\bS$ being $i$-informative for some $i \in I$ is at least
\[ \Omega\left( \dfrac{|I|}{\log^{2} n} \cdot \dfrac{\binom{n-1}{2^t-1}}{\binom{n}{2^t}}\right) = \Omega\left( \dfrac{\alpha}{\log^{2} n}\right). \]
This finishes the proof of the lemma.
\end{proof}

By our choice of $K$, a set $S$ that is $i$-informative for some $i\in I$ is sampled during the $K$ main loops with probability $1-o(1)$. By Lemma \ref{lem:case2-1} a violation is found with probability $1-o(1)$.

\subsection{Proof of Lemma \ref{informative-lemma}}\label{proof-informative}

\begin{proof}
Let $\gamma$ denote the fraction of $S\in \calP_{i,t}$ that
  are not informative for the $i$th coordinate. 
Then by definition, at least one of the two conditions must hold:
\begin{flushleft}\begin{enumerate}
\item At least $\gamma/2$-fraction of $S\in \calP_{i,t}$ have
  $\goodf_i^+(S)<0.1\cdot {\wteps^2}/({\alpha \sqrt{n}})$; or,\vspace{-0.06cm}
\item At least $\gamma/2$ fraction of $S\in \calP_{i,t}$ have
  at least $0.9$-fraction of $(2^r-1)$-sized subsets $T\subseteq S$ have
  $\goodf_i^-(T)<0.1\cdot {\wteps^2}/({\beta \sqrt{n}})$.
\end{enumerate}\end{flushleft}
Below we show that $\gamma\le 5/8$ in the first case and
  $\gamma\le 7/8$ in the second case.

We start with the first case, where at least $\gamma / 2$ fraction of $S \in \calP_{i, t}$ have 
$$\goodf_i^+(S) < 0.1 \cdot \frac{\wteps^2}{\alpha \sqrt{n}}.$$ Consider the following two methods of sampling a pair $(\bx, \bS)$ which is \emph{not good} for $E_i^+$:
\begin{flushleft}\begin{itemize}
\item We first sample $\bx$ from $\strong_i^+$ and then $\bS$ from $\calP_{i, t}$, both uniformly at random.\vspace{-0.13cm}
\item We first sample $\bS$ from $\calP_{i, t}$ and then $\bx$ from $\strong_i^+$, both uniformly at random.
\end{itemize}\end{flushleft}
The probabilities of sampling a pair $(\bx,\bS)$ that is not good for $E_i^+$ under the two methods are the same 
  since both are equal to the fraction of $(x,S)$ that are not good among $\strong_i^+\times \calP_{i,t}$.
Using the first way of sampling, we have that the probability that $(\bx,\bS)$ is not good is at most $1/4$,
since each $x \in \strong_i^+$ has at least $(3/4)$-fraction of $S \in \calP_{i, t}$ such that $(x, S)$ is a good pair. Using the second method, on the other hand, we note that
\[ \Prx \big[(\bx, \bS) \text{ is not good}\big] \geq \frac{\gamma}{2} \cdot \left(1 - \left(0.1\cdot \frac{\wteps^2}{\alpha \sqrt{n}}\right)\cdot 2^n\cdot \frac{1}{|\strong_i^+|} \right) \geq \dfrac{\gamma}{2} \cdot 0.8, \]
where we used $H\le 2|I|/\wteps^2$ and thus,
\[ \dfrac{|\strong_i^+|}{2^n} \geq \dfrac{\sqrt{n}}{H 2^t} \ge \frac{\sqrt{n}\hspace{0.04cm}\wteps^2}{2|I|2^t}= \dfrac{\wteps^2}{2\alpha \sqrt{n}}. \] 
Combining both inequalities, we obtain that $\gamma \leq  {5}/{8}$.

Next we consider the second case using a similar argument. We sample a pair $(\bx, \bT)$ which is not good for $E_i^-$ using the following two methods:
\begin{flushleft}\begin{itemize}
\item We first sample $\bx$ from $\strong_i^-$, $\bS$ from $\calP_{i, t}$, and then sample $\bT\subseteq \bS$ of size $2^r-1$, which is essentially sampling $\bT$ uniformly from $\calP_{i, r}$.\vspace{-0.06cm}
\item We first sample $\bS \in \calP_{i, t}$ uniformly at random, and then sample a subset $\bT \subseteq \bS$ of size $2^{r} - 1$ uniformly at random, and finally we sample $\bx$ from $\strong_i^-$. 
\end{itemize}\end{flushleft}
Similarly to the first case,  the probability of sampling a pair $(\bx,\bT)$ that is 
  not good is at most $1/4$ using the first method. 
Using the second method, we obtain a lower bound on the probability:
\[ \Prx \big[(\bx, \bT) \text{ is not good}\big] \geq \frac{\gamma}{2} \cdot 0.9 \cdot \left( 1 - 
\left( 0.1\cdot \frac{\wteps^2}{\beta \sqrt{n}}\right)\cdot 2^n\cdot \frac{1}{|\strong_i^-|}\right) \geq \frac{\gamma}{2} \cdot 0.9\cdot 0.8. \]
Combining the above two inequalities, we obtain that $\gamma \leq  {7}/{8}$. 
\end{proof}



\section{Proof of Lemma \ref{main2}}\label{sec:main2}

We first show that Lemma \ref{main2} follows from the following lemma that we prove in this section.

\begin{lemma}\label{main4}
If $f\colon\{0,1\}^n\rightarrow \{0,1\}$ is $\eps$-far from unate and $I_f \leq 6\sqrt{n}$,
  then we have
\begin{equation}\label{maineq4}
\sum_{i=1}^n \score_i^- \geq \Omega\left(\dfrac{\eps^2}{\log^8 n}\right).
\end{equation}
\end{lemma}
\begin{proof}[Proof of Lemma \ref{main2} assuming Lemma \ref{main4}]
Suppose that the LHS of (\ref{maineq}) is achieved by
$$
\sum_{i\in W}\score_i^+ +\sum_{i\notin W} \score_i^-
$$
for some $W\subseteq [n]$.
Then we let $r\in \{0,1\}^n$ be the string with $r_i=1$ if $i\in W$ and $r_i=0$ if $i\notin W$,
  and let $g$ be the Boolean function with $g(x)=f(x\oplus r)$.
On the one hand, $g$ has the same distance to unateness as $f$ and satisfies $I_g=I_f$,
  so Lemma \ref{main4} applies to $g$.
On the other hand, it follows from the description of $\ESearch$ and
  the definition of $\score$ that $\score_i^-$ of $g$ is exactly the same
  as $\score_i^+$ of $f$ if $i\in W$ or $\score_i^-$ of $f$ if $i\notin W$.
To see this is the case, note that the output distribution of $\ESearch(x,S)$ on $g$
  is exactly the same as that of $\ESearch(x\oplus r,S)$ on $f$.
As a result, whether $(x,S)$ is a good pair
  or not in $g$ is the same as that of $(x\oplus r,S)$ in $f$ (except that the roles of $E_i^+$ and $E_i^-$
  may get switched depending on whether $i\in W$ or not).
(\ref{maineq}) for $f$ then follows from (\ref{maineq4}) for $g$.
\end{proof}

We prove Lemma \ref{main4} in the rest of the section.
Let $f$ be a function that is $\eps$-far from unate and has $I_f = O(\sqrt{n})$.
Let $G_f$ be the bipartite graph of \emph{anti-monotone edges} of $f$ defined as follows:
\begin{flushleft}\begin{enumerate}
\item Vertices on the LHS of $G_f$ correspond to points $x\in \{0,1\}^n$ with $f(x)=1$ and vertices
  on the RHS of $G_f$ correspond to points $y\in \{0,1\}^n$ with $f(y)=0$;
\item $(x,y)$ is an edge in $G_f$ if $(x,y)$ is an anti-monotone edge.
\end{enumerate}\end{flushleft}
We recall a key technical lemma from \cite{KMS15} which states that when $f$ is $\eps$-far from
  monotone (which is the case here since $f$ is $\eps$-far from unate), $G_f$ must contain a large and ``good'' subgraph.

\begin{definition}
Let $G=(U,V,E)$ be a bipartite subgraph of $G_f$, where $U$ is a set of points $x$ with $f(x)=1$, $V$ is a set of points $y$ with $f(y)=0$, and $E$ consists
  of anti-monotone edges of $f$.
We say $G$ is right-$d$-good, for some positive integer $d$, if the degree of every $y\in V$ is in the range $[d,2d]$ and
  the degree of every $x\in U$ is at most $2d$, and $G$ is left-$d$-good if the degree
  of every $x\in U$ is in the range $[d,2d]$ and the degree of every $y\in V$ is at most $2d$.
\end{definition}

\begin{theorem}[Lemma 7.1 in \cite{KMS15}]
\label{thm:kms}
If $f$ is $\eps$-far from monotone, then $G_f$ contains a bipartite subgraph $G=(U,V,E)$
  that satisfies one of the following conditions:
\begin{enumerate}
\item $G$ is left-$d$-good for some positive integer $d$ and
  $\sigma=|U|/2^n$ satisfies
\begin{equation}\label{hehehehe}
\sigma^2 d= \Theta\left(\frac{\eps^2}{\log^4 n}\right).
\end{equation}
\item $G$ is right-$d$-good for some positive integer $d$ and
  $\sigma=|V|/2^n$ satisfies (\ref{hehehehe}).
\end{enumerate}
\end{theorem}

Since $f$ is $\eps$-far from unate and in particular, $\eps$-far from monotone,
  Theorem \ref{thm:kms} applies to $f$ and we use $G=(U,V,E)$ to denote such a subgraph of $G_f$.
In the rest of the proof we assume without loss of generality that
  $G$ is left-$d$-good and $\sigma=|U|/2^n$ satisfies (\ref{hehehehe});
  the proof for the other case when $G$ is right-$d$-good is symmetric.
Given $G$ and $\sigma$, we choose the $t$ to be the largest integer with
\begin{equation}\label{choiceoft}
2^{t-1}\le \left\lceil \frac{\sigma\sqrt{n}}{\log^4 n} \right\rceil.
\end{equation}
So we have $t\ge 1$. Using $\sigma\le 1$, we also have $t< \Lambda$ and thus, $t\in [\Lambda]$.

Our goal is to show the following lemma from which Lemma \ref{main4} follows directly:
\begin{lemma}\label{main3}
Let $G = (U, V, E)$ be a subgraph of $G_f$ that is left-$d$-good and satisfies (\ref{hehehehe}).
Then 
\begin{equation} \dfrac{2^t}{\sqrt{n}}\cdot \sum_{i=1}^n \score_{i, t}^{-} = \Omega\left(\dfrac{\eps^2}{\log^8 n}\right) . \label{eq:maineq22}
\end{equation}
\end{lemma}

To gain some intuition, assume
  that every $(x,y)\in E$ in direction $i$ has $x$ being $t$-strong for $E_i^-$ (we always use
  $x$ to denote points in $U$ and $y$ or $x^{(i)}$ to denote points in $V$ in the rest of the proof). Then, each edge
  in $E$ contributes ${1}/{2^n}$ to the sum on the LHS of (\ref{eq:maineq22}) and thus,
  it is at least
\begin{equation}
\frac{2^t}{\sqrt{n}}\cdot \frac{|E|}{2^n}\ge \frac{2^t}{\sqrt{n}}\cdot \frac{d|U|}{2^n}=
\dfrac{2^t}{\sqrt{n}} \cdot \sigma d = \Omega\left( \dfrac{\sigma \sqrt{n}}{\log^4 n \cdot \sqrt{n}} \right) \cdot \sigma d = \Omega\left( \dfrac{\sigma^2 d}{\log^4 n} \right) = \Omega\left(\dfrac{\eps^2}{\log^8 n} \right),\label{eq:deduce}
\end{equation}
where the first inequality follows from the fact that $G$ is left-$d$-good
  and the last equation uses (\ref{hehehehe}). Note that for the specific case when $t=1$, every edge $(x, y) \in E$ in direction $i$ is $t$-strong for $E_i^{-}$; thus, this intuition holds formally, and we may assume for the rest of the proof that $t > 1$.

The plan for the rest of the proof is as follows. First we give
  a sufficient condition for a point $x\in U$ to be $t$-strong for $E_i^-$.
Next, we prove a persistency lemma similar to that from \cite{KMS15} to show that for
  most edges $(x,x^{(i)})\in E$, $x$ is $t$-strong
  for $E_i^-$. 
Finally, by a similar argument to (\ref{eq:deduce}), we obtain the same conclusion only losing only a constant factor in the inequality.

More specifically, in Section~\ref{sec:robust} we define the notion of \emph{robust} sets, and show that if an edge
  $(x,x^{(i)})\in E$
has many robust sets, its left endpoint $x$ is $t$-strong for $E_i^-$.
Then in Section~\ref{sec:solid}, we introduce the notion of \emph{solid} edges, 
  and show that 1) most edges of $E$ are solid and 2) every {solid} edge have many robust sets, which then finishes the proof.

\subsection{Edges with many robust sets have strong left endpoints}\label{sec:robust}

We now introduce the notion of robust sets $S\in \calP_{i,t}$ for anti-monotone edges $(x,x^{(i)})\in E$
  (with $x\in U$ being the left endpoint and $i$ being the direction).
Intuitively, we would like to show that if
  $S\in \calP_{i,t}$ is a robust set for $(x,x^{(i)})$,
  then $\ESearch(x, S \cup \{ i \})$ returns $i$ with probability at least $\frac{1}{2}$ and thus,
  $(x,S)$ is a good pair for $E_i^-$.
This is shown in Lemma \ref{lem:robust}.
As a result, to show that $x$ is $t$-strong for $E_i^-$, it suffices to show
  that most sets $S\in \calP_{i,t}$ are robust for $(x,x^{(i)})$.

\begin{definition}[Robust Sets]\label{def:robust}
Let $(x, x^{(i)})\in E$. 
We say $S\in \calP_{i,t}$ is a \emph{robust} set for $(x, x^{(i)})$ if:
\begin{enumerate}
\item At least $(1 - ( {1}/{\log^2 n}))$-fraction of subsets $S' \subset S$ of size $2^{t-1} - 1$ has $f(x^{(S' \cup \{ i \})}) \neq f(x)$.\vspace{-0.1cm}
\item At least $(1 - ( {1}/{\log^2 n}))$-fraction of subsets $S' \subset S$ of size $2^{t-1}$ has $f(x^{(S')}) = f(x)$.
\end{enumerate}
Note that for the special case when $t=1$ (letting $S=\{j\}$ for some $j\ne i$),
  conditions 2 and 3 above require
  $f(x^{(i)})\ne f(x)$ and $f(x^{(j)})=f(x)$, respectively.
\end{definition}

We show that if $S\in \calP_{i,t}$ is robust for an edge $(x,x^{(i)})\in E$, then
  $(x,S)$ is a good pair for $E_i^-$.

\begin{lemma}\label{lem:robust}
If $S\in \calP_{i,t}$ is a robust set for $(x,x^{(i)})\in E$, then $\ESearch(x, S\cup\{i\})$
  returns $i$ with high probability (and thus, $(x,S)$ is a good pair for $E_i^-$).
\end{lemma}
\begin{proof}
Let $S\in \calP_{i,t}$ be a robust set for $(x,x^{(i)})\in E$.
The proof consists of two claims.
The first claim shows that in $\ESearch(x,S\cup\{i\})$,
  we have $i\in C$ at the end of Step 3 with high probability.
The second claim shows that $C$ does not contain any $\ell\in S$ with high probability.


\begin{claim}
With probability at least $1 - o(1)$, we have $i \in \bC$ in Step $3$ of $\ESearch(x,S\cup \{i\})$.
\end{claim}
\begin{proof}
We use $\bT_1, \ldots, \bT_L$ to denote the $L$ random subsets sampled in Step 3 of $\ESearch$.
For $i\notin \bC$ to happen, either every $\bT_\ell$ satisfies $f(x^{(\bT_\ell)})=f(x)$,
  or one of the $\bT_\ell$ satisfies $f(x^{(\bT_\ell)})\ne f(x)$
  and $i\notin \bT_\ell$. 
Below we upperbound the probability of each of these two events by $o(1)$.
The claim then follows by a union bound on the two events.

First, by condition 1 of Definition \ref{def:robust},
  the probability of $f(x^{(\bT_\ell)})\ne f(x)$ for each $\ell$ is at least (by only considering the case when $i\in \bT_\ell$)
$$
(1/2)\cdot (1-(1/\log^2 n))
$$
and thus, the probability of the first event is $o(1)$ using our choice of $L=\lceil 4\log n\rceil$.


Next by condition 2 of Definition \ref{def:robust},
  the probability of $i\notin \bT_\ell$ and $f(x^{(\bT_\ell)})\ne f(x)$ is at most
$$
(1/2)\cdot (1/\log^2 n)=1/(2\log^2 n).
$$
By a union bound on $\ell \in [L]$ and our choice of $L$,
  the probability of the second event is also $o(1)$.
\end{proof}

\begin{claim}
With probability at least $1-o(1)$, we have $\bC \cap S=\emptyset$ in Step 3 of $\ESearch(x,S\cup\{i\})$.
\end{claim}
\begin{proof}
Consider an index $k \in S$ and note that $k \neq i$. Then, in order to have $k\in \bC$,
  none of the $\bT_\ell$ can satisfy both $f(x^{(\bT_\ell)})\ne f(x)$ and $k\notin \bT_\ell$.
However,
there are $\smash{\binom{2^t-2}{2^{t-1} - 1}}$ subsets of $S \cup \{ i\}$ of size $2^{t-1}$ which include $i$ and exclude $k$. Among them the number of $T$ with $f(x^{(T)})=f(x)$ is at most $$\frac{1}{\log^2 n} \cdot \binom{2^{t} - 1}{2^{t-1} - 1}$$
by condition 2. So the fraction of $T$ that includes $i$, excludes $k$, and has $f(x^{(T)})\ne f(x)$ is at least
\[\dfrac{\binom{2^t-2}{2^{t-1} - 1} - \frac{1}{\log^2 n} \cdot \binom{2^{t} - 1}{2^{t-1} - 1}}{\binom{2^t}{2^{t-1}}} \geq \frac{1}{4} - \frac{1}{2\log^2 n}=\frac{1}{4}-o(1). \]
Thus, the probability of $k \in \bC$ is at most $(3/4+o(1))^L$.
This is $o(1/n)$ by our choice of $L=\lceil 4\log n\rceil $.
The lemma then follows by a union bound over all $k\in S$.
\end{proof}

It follows that $\bC=\{i\}$ with probability $1-o(1)$.
This finishes the proof of Lemma \ref{lem:robust}.
\end{proof}

\subsection{Solid edges have many robust sets}\label{sec:solid}

We now introduce the notion of $(\tau,\gamma)$-persistent.
These are points $x$ at which the value of $f$ remains the same as $f(x)$ with high probability (at least $1-\gamma$)
  after flipping $\tau$ random bits of $x$.

\begin{definition}[Persistent Points]
Let $\tau:0\le \tau\le n$ be a nonnegative integer and $\gamma\in [0,1]$.
We say a point $x\in \{0,1\}^n$ is
  $(\tau,\gamma)$-persistent in $f$ if, by flipping a set $\bS\subseteq [n]$ of size $\tau$ drawn uniformly at random, we have
  $\Pr_{\bS} [f(x)\ne f(x^{(\bS)}) ]\le \gamma.$
For the special case of $\tau=0$, every $x\in \{0,1\}^n$ is $(0,\gamma)$-persistent for all $\gamma\in [0,1]$.
\end{definition}
\begin{remark}
\begin{flushleft}For readers familiar with \emph{\cite{KMS15}},
  our definition of persistency is slightly different from that of \emph{\cite{KMS15}}: 1) we need the second parameter
  $\gamma$ while \emph{\cite{KMS15}} always sets $\gamma=1/10$;
  2) More importantly, \emph{\cite{KMS15}} only allows one to randomly flip $0$-entries of $x$ to $1$
  (since they are interested in monotonicity testing) while we flip all sets of $\tau$ coordinates
  of $x$ uniformly at random.
\end{flushleft}\end{remark}

We need the following persistency lemma. Its proof is similar to Lemma 9.3 of \cite{KMS15}.
Due to differences discussed above, we give a self-contained proof adapted from \cite{KMS15}
  in Appendix~\ref{persistency-proof}.


\begin{lemma}[Persistency lemma]\label{persistency-lemma}
For $\gamma\in(0,1]$ and $\tau\in[0,n]$, the fraction of $(\tau,\gamma)$-non-persistent points is 
  at most $O(I_f\cdot \tau/(n\gamma))$.
\end{lemma}

Let $\gamma=1/{\log^3 n}$. As $I_f=O(\sqrt{n})$, the fraction of $(\tau,\gamma)$-non-persistent
  points is at most $$O\left(\frac{\tau \cdot \log^3 n}{\sqrt{n}}\right).$$
Next we introduce the notion of \emph{solid} edges in $G$ and prove using the persistency lemma that
  most edges in $G$ are solid.

\begin{definition}[Solid Edges]
An edge $(x,y)$ in $G$ (with $x\in U$ and $y\in V$ as usual) is \emph{solid} if $x$ is $(2^{t-1},\gamma)$-persistent
  and $y$ is $(2^{t-1} - 1,\gamma)$-persistent.
\end{definition}


\begin{lemma}\label{manysolid}
At least $(1-o(1))$-fraction of edges of $G$ are solid.
\end{lemma}
\begin{proof}
First recall that the number of edges in $G$ is $\Theta(\sigma 2^n\cdot d)$.
Using the persistency lemma,
  the fraction of points (with respect to the full set of size $2^n$)
  that are $(2^{t-1},\gamma)$-non-persistent is at most
\begin{equation}\label{hehehehehe}
O\left(\frac{2^{t-1} \cdot \log^3 n}{\sqrt{n}}\right). 
\end{equation}
By our choice of $t$ in (\ref{choiceoft}), and the fact that $t > 1$,
$\sigma\sqrt{n}/\log^4 n\ge 1$, in which case we have $2^{t-1}=O(\sigma\sqrt{n}/\log^4 n)$
  and (\ref{hehehehehe}) becomes $O(\sigma/\log n)=o(\sigma)$,

As a result, the number of edges $(x,y)$ in $G$ with $x$ being $(2^{t-1},\gamma)$-non-persistent
  is $o(\sigma 2^n\cdot d)$.
By a similar argument one can show that the number of edges $(x,y)$ in $G$ with $y$
  being either $(2^{t-1}-1,\gamma)$-non-persistent is
  $o(\sigma2^n \cdot d)$, at most a $o(1)$-fraction of edges in $G$.
\end{proof}


Finally we show that every solid edge in $G$ has many robust sets in $\calP_{i,t}$.

\begin{lemma}\label{manyrobust}
Suppose that $(x, x^{(i)})\in E$ is a solid edge in $G$. Then the fraction of sets $S\in \calP_{i,t}$
  that are not robust with respect to edge $(x,x^{(i)})$ is at most ${4}/{\log n}.$
\end{lemma}

\begin{proof}
By definition, a set $S\in \calP_{i,t}$ is not robust for $(x,x^{(i)})$
  if one of the following events occur:
\begin{enumerate}
\item At least $(1/\log^2 n)$-fraction of subsets $T\subset S$ of size $2^{t-1} - 1$
  have $f(x^{(T \cup \{i \})}) = f(x)\ne f(x^{(i)})$.\vspace{-0.12cm}
\item At least $(1/\log^2 n)$-fraction of subsets $T\subset S$ of size $2^{{t-1}}$
  satisfy $f(x^{(T)}) \neq f(x)$.
\end{enumerate}
We bound the fraction of such $S$ in $\calP_{i,t}$ for each event separately.

For the second event, we consider the following bipartite graph $H'$.
Vertices $U'$ on the LHS of $H'$ correspond to subsets $T$ of $[n]\setminus \{i\}$ of size
  $2^{t-1}$; vertices on the RHS correspond to $S\in \calP_{i,t}$;
  $(T,S)$ is an edge iff $T\subset S$.
Clearly $H'$ is bi-regular. Let $d_{\text{left}}$ and $d_{\text{right}}$ denote the degrees.
Let $\alpha$ be the fraction of $S$ among $\calP_{i,t}$ (the RHS) for which the second event occurs.
Then the number of $T$ with $f(x^{(T)})\neq f(x)$ is at least
$$
\alpha \cdot |\calP_{i,t}|\cdot d_{\text{right}} \cdot \frac{1}{\log^2 n} \cdot\frac{1}{d_{\text{left}}}
=\alpha\cdot |U'| \cdot \frac{1}{\log^2 n} =\alpha \cdot \frac{1}{\log^2 n} \cdot \binom{n-1}{2^{t-1}}.
$$
However, as $x$ is $(2^{t-1},\gamma)$-persistent, the number of such $T$ is
  at most $
\gamma \cdot \binom{n}{2^{t-1}}.
$
As a result we have
$$
\alpha \le \log^2 n \cdot \gamma \cdot \frac{\binom{n}{2^{t-1}}}{\binom{n-1}{2^{t-1}}} \le \frac{2}{\log n}.
$$

For the first event, we consider a similar regular bipartite graph $H^*$. Vertices
  $U^*$ on the LHS correspond to subsets $T$ if $[n] \setminus \{ i \}$ of size $2^{t-1} -1$; vertices on the RHS correspond to $S \in \calP_{i,t}$; $(T,S)$ is an edge iff $T \subset S$. Similarly, we let $\beta$ be the fraction of $S$
    among $\calP_{i,t}$ (the RHS) for which the first event occurs. Then the number of $T$ with $f(x^{(T \cup \{ i \})}) =f(x)\ne f(x^{(i)})$ is at least
\[ \beta \cdot |\calP_{i,t}| \cdot d_{\text{right}} \cdot \dfrac{1}{\log^2 n} \cdot \frac{1}{d_\text{left}} = \beta \cdot |U^*| \cdot \dfrac{1}{\log^2 n} = \beta \cdot \frac{1}{\log^2 n} \cdot \binom{n - 1}{2^{t-1}-1}. \]
However, since $x^{(i)}$ is $(2^{t-1}-1, \gamma)$-persistent, the number of such $T$ is at most
$ \gamma \cdot \binom{n}{2^{t-1} - 1} $.
Thus,
\[ \beta \leq \log^2n \cdot \gamma\cdot \dfrac{ \binom{n}{2^{t-1}-1}}{\binom{n-1}{2^{t-1}-1}} \leq \frac{2}{\log n}. \]
So the total fraction of sets $S\in \calP_{i,t}$ that satisfy one of the events is at most ${4}/{\log n}$.
\end{proof}

\subsection{Finishing the proof of Lemma \ref{main3}}

From Lemma \ref{manysolid}, we know the number of solid edges in $G$ is at least $\Omega(|E|)$.
Combining Lemma \ref{manyrobust} with Lemma \ref{lem:robust}, we know for each solid edge $(x,x^{(i)})$ in $G$, at least $(1-4/\log n)$-fraction of sets in $\calP_{i,t}$ are robust, and its left end point
  $x$ is $t$-strong for $E_i^-$ and contributes $1/2^n$ to
  $\score_{i,t}^-$ in the LHS of (\ref{eq:maineq22}).
Lemma \ref{main3} then follows from the same analysis done in (\ref{eq:deduce}), after
  replacing $|E|$ by $\Omega(|E|)$ at the beginning.

\section*{Acknowledgments}

We thank Rocco Servedio and Li-Yang Tan for countless discussions and suggestions.
This work~is supported in part by NSF CCF-1149257, NSF CCF-142310, and a
  NSF Graduate Research Fellowship under Grant No. DGE-16-44869.

\begin{flushleft}
\bibliographystyle{alpha}
\bibliography{waingarten}

\newcommand{\etalchar}[1]{$^{#1}$}
\begin{thebibliography}{BCP{\etalchar{+}}17b}

\bibitem[BB16]{BB16}
Aleksandrs Belovs and Eric Blais.
\newblock A polynomial lower bound for testing monotonicity.
\newblock In {\em Proceedings of the 48th {ACM} Symposium on the Theory of
  Computing ({STOC}~'2016)}, pages 1021--1032, 2016.

\bibitem[BCP{\etalchar{+}}17a]{BCPRS17b}
Roksana Baleshzar, Deeparnab Chakrabarty, Ramesh Krishnan~S. Pallavoor, Sofya
  Raskhodnikova, and C.~Seshadhri.
\newblock A lower bound for nonadaptive, one-sided error testing of unateness
  of boolean functions over the hypercube.
\newblock {\em arXiv preprint arXiv:1706.00053}, 2017.

\bibitem[BCP{\etalchar{+}}17b]{BCPRS17}
Roksana Baleshzar, Deeparnab Chakrabarty, Ramesh Krishnan~S. Pallavoor, Sofya
  Raskhodnikova, and C.~Seshadhri.
\newblock Optimal unateness testers for real-values functions: Adaptivity
  helps.
\newblock In {\em Proceedings of the 44th International Colloquium on Automata,
  Languages and Programming ({ICALP}~'2017)}, 2017.

\bibitem[Bla09]{B09}
Eric Blais.
\newblock Testing juntas nearly optimally.
\newblock In {\em Proceedings of the 41st {ACM} Symposium on the Theory of
  Computing ({STOC}~'2009)}, pages 151--158, 2009.

\bibitem[BMPR16]{BMPR16}
Roksana Baleshzar, Meiram Murzabulatov, Ramesh Krishnan~S. Pallavoor, and Sofya
  Raskhodnikova.
\newblock Testing unateness of real-valued functions.
\newblock {\em arXiv preprint arXiv:1608.07652}, 2016.

\bibitem[CC16]{CS16}
Deeparnab Chakrabarty and Seshadhri Comandur.
\newblock An o(n) monotonicity tester for boolean functions over the hypercube.
\newblock {\em {SIAM} Journal on Computing}, 45(2):461--472, 2016.

\bibitem[CDST15]{CDST15}
Xi~Chen, Anindya De, Rocco~A. Servedio, and Li-Yang Tan.
\newblock Boolean function monotonicity testing requires (almost) {$n^{1/2}$}
  non-adaptive queries.
\newblock In {\em Proceedings of the 47th {ACM} Symposium on the Theory of
  Computing ({STOC}~'2015)}, pages 519--528, 2015.

\bibitem[CG17]{CG17}
Cl\'{e}ment~L. Canonne and Tom Gur.
\newblock An adaptivity hierarchy theorem for property testing.
\newblock {\em arXiv preprint arXiv:1702.05678}, 2017.

\bibitem[CS14]{CS14}
Deeparnab Chakrabarty and C.~Seshadhri.
\newblock An optimal lower bound for monotonicity testing over hypergrids.
\newblock {\em Theory of Computing}, 10(17):453--464, 2014.

\bibitem[CS16]{CS16b}
Deeparnab Chakrabarty and C.~Seshadhri.
\newblock A {$\widetilde{O}(n)$} non-adaptive tester for unateness.
\newblock {\em arXiv preprint arXiv:1608.06980}, 2016.

\bibitem[CST14]{CST14}
Xi~Chen, Rocco~A. Servedio, and Li-Yan Tan.
\newblock New algorithms and lower bounds for monotonicity testing.
\newblock In {\em Proceedings of the 55th Annual {IEEE} Symposium on
  Foundations of Computer Science ({FOCS}~'2014)}, pages 285--295, 2014.

\bibitem[CWX17]{CWX17}
Xi~Chen, Erik Waingarten, and Jinyu Xie.
\newblock Beyond talagrand functions: new lower bounds for testing monotonicity
  and unateness.
\newblock In {\em Proceedings of the 49th {ACM} Symposium on the Theory of
  Computing ({STOC}~'2017)}, 2017.

\bibitem[GGL{\etalchar{+}}00]{GGLRS00}
Oded Goldreich, Shafi Goldwasser, Eric Lehman, Dana Ron, and Alex Samordinsky.
\newblock Testing monotonicity.
\newblock {\em Combinatorica}, 20(3):301--337, 2000.

\bibitem[KMS15]{KMS15}
Subhash Khot, Dor Minzer, and Muli Safra.
\newblock On monotonicity testing and boolean isoperimetric type theorems.
\newblock In {\em Proceedings of the 56th Annual {IEEE} Symposium on
  Foundations of Computer Science ({FOCS}~'2015)}, pages 52--58. IEEE Computer
  Society, 2015.

\bibitem[KS16]{KS16}
Subhash Khot and Igor Shinkar.
\newblock An {$\widetilde{O}(n)$} queries adaptive tester for unateness.
\newblock In {\em Approximation, Randomization and Combinatorial Optimization.
  Algorithms and Techniques}, pages 37:1--37:7, 2016.

\bibitem[Tal93]{T93}
Michel Talagrand.
\newblock Isoperimetry, logarithmic sobolev inequalities on the discrete cube,
  and margulis' graph connectivity theorem.
\newblock {\em Geometric {\&} Functional Analysis}, 3(3):295--314, 1993.

\end{thebibliography}
\end{flushleft}

\appendix

\section{Proof of the persistency lemma}\label{persistency-proof}

We give a proof of Lemma \ref{persistency-lemma}, which is a slight modification of Lemma 9.3 of \cite{KMS15}.

\begin{lemma}[Persistency lemma]
For $\gamma\in(0,1]$ and $\tau\in[0,n]$, the fraction of $(\tau,\gamma)$-non-persistent points is at most $O(I_f\cdot \tau/(n\gamma))$.
\end{lemma}

\begin{proof}
Let $\eta$ be the fraction of $(\tau, \gamma)$-non-persistent points. Consider the following random process:
\begin{enumerate}
\item we draw $\bx\in\{0,1\}^n$ uniformly at random.\vspace{-0.12cm}
\item we flip $\tau$ bits from $\bx$ uniformly at random to get $\by \in \{0, 1\}^n$.
\end{enumerate}
Let $(\bx, \by) \sim \calD$ be the distribution supported on pairs $x, y \in \{0, 1\}^n$ given by the above procedure. Let $\calD_{x}$ be the distribution supported on $\{0, 1\}^n$ given by sampling $\by$ where $(\bx, \by) \sim \calD$ conditioned on $\bx = x$.
For each $x\in\{0,1\}^n$, let
\[ r_{x} =\begin{cases} \Prx_{\by \sim \calD_{x}}[f(x) \neq f(\by)] &  \text{if $x$ is $(\tau,\gamma)$-non-persistent} \\
							0 & \text{otherwise} \end{cases} \]

By definition, for $(\tau,\gamma)$-non-persistent points $x$ we have $r_x\geq \gamma$, which implies:

\[ \Prx_{(\bx, \by) \sim \calD}[f(\bx) \neq f(\by)] = \frac{1}{2^n} \sum_{x \in \{0, 1\}^n} \Prx_{\by \sim \calD_{x}}[f(x) \neq f(\by)] \geq \frac{1}{2^n} \sum_{x \in \{0, 1\}^n} r_{x} \geq \gamma \cdot \eta \]

On the other hand, consider the following random process:
\begin{enumerate}
\item We draw $\bx_0 \in \{0, 1\}^n$ uniformly at random.\vspace{-0.12cm}
\item For each $k \in [\tau]$,  pick a random coordinate $\bi \in [n]$ that has not been chosen and let $\bx_k = \bx_{k-1}^{(\bi)}$.
\end{enumerate}
Note that $(\bx_0, \bx_{\tau})$ is distributed as $\calD$. Additionally, we have
\[ \Prx_{(\bx, \by) \sim \calD}[f(\bx) \neq f(\by)] \leq \sum_{k=1}^{\tau} \Prx_{\bx_{k-1}, \bx_{k}}\left[f(\bx_{k-1}) \neq f(\bx_{k}) \right] \leq \tau \cdot \frac{I_f}{n}, \]
where in the last step, we used the fact that each $(\bx_{k-1}, \bx_{k})$ is distributed as a uniform edge of the hypercube.
Combining the two inequalities above, we obtain $\eta = O ( (I_f \cdot \tau)/{(n\cdot \gamma)} )$.
\end{proof}

\section{Unateness testing algorithm for high-influence functions}
\label{sec:high-inf}

We note that implicitly, Lemma 9.1 of \cite{KMS15} proves the following lemma:
\begin{lemma}\label{blabla1}
If $f \colon \{0, 1\}^n \to \{0, 1\}$ has $I_f > 6 \sqrt{n}$, then
\begin{equation}\label{lalahehe}
\frac{1}{2^n} \sum_{i=1}^n \min \big\{ |E_i^+|, |E_i^-| \big\} \geq 2 \sqrt{n}.
\end{equation}
\end{lemma}

In particular, it implies that if $I_f\ge 6\sqrt{n}$ then $f$ cannot be unate.
Next we slightly modify the algorithm of \cite{CS16b} to obtain the following lemma:
\begin{lemma}\label{blabla2}
There is an $ O\ ( \sqrt{n} \cdot \log^2(n) )$-query, non-adaptive algorithm that, given
  any function $f$ that satisfies (\ref{lalahehe}),
  finds an edge violation of $f$ to unateness with probability at least $2/3$.
\end{lemma}

Lemma \ref{hehelem} follows by combining Lemma \ref{blabla1} and Lemma \ref{blabla2}. For completeness, we include the algorithm of \cite{CS16b} (with modified parameters) in order to prove Lemma~\ref{blabla2}.

\begin{figure}[t!]
\begin{framed}
\noindent \texttt{Unate-E-Tester}, the unateness tester from \cite{CS16b} which rejects functions satisfying (\ref{lalahehe}).
\vspace{-0.16cm}
\begin{flushleft}\noindent {\bf Input:} Query access to $f \colon \{0, 1\}^n \to \{0, 1\}$ satisfying (\ref{lalahehe}).

\noindent {\bf Output:}  Either ``unate,'' or two edges constituting an edge violation for $f$.

\begin{itemize}

\item[] \vskip -.02in \hspace{-0.2cm}Repeat the following for $r = 1, 2,\dots, L = \lceil \log(8n) \rceil$: \vspace{-0.08cm}

\begin{enumerate}

\item \vskip -.02in Repeat the following $s_r = \left\lceil \dfrac{20 \sqrt{n}\cdot \log(8n)}{2^r} \right\rceil$ times: 

\item  Sample a direction $\bi \sim [n]$ uniformly at random and sample $4 \cdot 2^r$ random edges in direction $\bi$ and query all end points. Return an edge violation if one is found.

\end{enumerate}
\item[] \vskip -0.045in \hspace{-0.2cm}If we have not found any violation in line 2, return ``unate.'' 
\end{itemize}
\end{flushleft}\vskip -0.14in
\end{framed}
\caption{Description of the \algsecond~for Case 2 of the algorithm.} \label{fig:case2}
\end{figure}

\begin{proof}[Proof of Lemma~\ref{blabla2}]
Let $\mu_i = \frac{\min\{|E_i^+|, |E_i^-|\}}{2^n}$. For any $r = [L]$, let $i \in S_r$ be the set of directions where 
\[ \frac{1}{2^r} \leq \mu_i \leq \frac{1}{2^{r-1}}. \]
If $\mu_i \leq \frac{1}{8n}$, then all such directions $i$ contribute at most $\frac{1}{8}$ to the RHS of (\ref{lalahehe}).
Thus, there exists one direction $r^*$ such that
\[ \frac{1}{2^n} \sum_{i \in S_{r^*}} \min\{ |E_i^+|, |E_i^-| \} \geq \frac{2\sqrt{n} - 1/8}{\log (8n)} \geq \frac{\sqrt{n}}{\log(8n)}, \]
and $|S_{r^*}| \geq \frac{2^{r^*} \sqrt{n}}{2\log(8n)}$. Thus, at the iteration corresponding to $r^*$ of \texttt{Unate-E-Tester}, the probability $\bi \in S_{r^*}$ is at least $\frac{2^{r^*}}{2\sqrt{n} \log(8n)}$. Conditioned on having sampled some $i \in S_{r^*}$, the probability that we do not observe an edge in $E_i^+$ in the first $2 \cdot 2^{r^*}$ edges is at most $(1 - \mu_i)^{2 \cdot 2^{r^*}} \leq e^{-2}$. Likewise, the probability we do not observe an edge in $E_i^-$ in the second $2 \cdot 2^{r^*}$ edges is at most $e^{-2}$, thus, with probability $1 - 2e^{-2} \geq \frac{7}{10}$, we observe an edge violation. 

Putting things together, the probability that we observe an edge violation in one iteration of line 2 with $r = r^*$ is at least $\frac{7 \cdot 2^{r^*}}{20 \cdot \sqrt{n} \log(8n)}$. Since we repeat line 2 for $s_r = \lceil \frac{20 \sqrt{n} \log (8n)}{2^{r^*}}\rceil$ times, we find an edge violation with high constant probability.
\end{proof}

\section{Overlap of two random sets of certain size}
\label{apphehe}

Let $k,\ell\in [n]$ be two positive integers with $\alpha=k\ell/n$.
We are interested in the size of $|\bS\cap \bT|$ where $\bS$ is a random $k$-sized subset of $[n]$
  and $\bT$ is a random $\ell$-sized subset of $[n]$, both drawn uniformly.

%

\begin{lemma}\label{appendix1}
For any $t\ge 4\alpha$, the probability of $|\bS\cap\bT|\ge  t$ is at most $\exp(-\Omega(t))$.
\end{lemma}
\begin{proof}
We assume without loss of generality that $k\ge \ell$.
If $\ell> n/2$, the claim is trivial as $\alpha> n/4$ and $t\ge 4\alpha> n$.
We assume $\ell\le n/2$ below.

We consider the following~process.
We draw $\bS$ first. Then we add random (and distinct) indices of $[n]$ to  $\bT$ round by round for $\ell$ rounds.
In each round we pick an index uniformly at random from those that
  have not been added to $\bT$ yet.
Clearly this process generates the same distribution of $\bS$ and $\bT$ that we are interested in.


For each $i\in [\ell]$, we let $\bX_i$ be the random variable that is set to $1$ if
  the index in the $i$th round belongs to $\bS$ and is $0$ otherwise.
Although $\bX_i$'s are not independent, the probability~of $\bX_i=1$ is at most
  $k/(n-\ell)\le 2k/n$ using $\ell\le n/2$,
  for any fixed values of $\bX_1,\ldots,\bX_{i-1}$.
Thus, the expectation of $\sum_{i\in [\ell]} \bX_i$ is at most $2k\ell/n=2\alpha$.
The lemma follows directly from the Chernoff bound (together with a standard coupling argument).
\end{proof}

\end{document}